\documentclass[12pt]{article}
\usepackage{amssymb,epsfig}
\newcommand{\idty}{{\leavevmode{\rm 1\mkern -5.4mu I}}}
\newcommand{\be}{\begin{eqnarray}}
\newcommand{\ee}{\end{eqnarray}}
\newcommand{\bea}{\begin{eqnarray*}}
\newcommand{\eea}{\end{eqnarray*}}
\renewcommand{\O}{\Omega}
\newcommand{\A}{{\cal A}}
\newcommand{\M}{{\cal M}}
\newcommand{\R}{\mathfrak{R}}
\renewcommand{\H}{\mathbb{H}}
\newcommand{\U}{\mathbb{U}}
\newcommand{\X}{\mathfrak{X}}
\renewcommand{\d}{\mbox{d}}
\newcommand{\oa}{\otimes_{\cal A}}
\newcommand{\pa}{\partial}
\renewcommand{\th}{\vartheta}
\newcommand{\ph}{\varphi}
\begin{document}
\renewcommand{\theequation} {\arabic{section}.\arabic{equation}}

\centerline{\huge \bf  Discrete Riemannian Geometry}
\vskip1cm
\begin{center} 
      {\bf A. DIMAKIS$^{1,2}$} and {\bf F. M\"ULLER-HOISSEN$^2$} 
       \vskip.5cm 
      \begin{minipage}{15cm}  \small
      $^1$Department of Mathematics, University of the Aegean,
      GR-83200 Karlovasi, Samos, Greece
      \vskip.1cm 
      \noindent
      $^2$Max-Planck-Institut f\"ur Str\"omungsforschung,
      Bunsenstrasse 10, D-37073 G\"ottingen, Germany     
      \end{minipage}
\end{center} 
\vskip.5cm

\abstract{Within a framework of noncommutative geometry, we develop an analogue
of (pseudo) Riemannian geometry on finite and discrete sets. On a finite set, there is a counterpart
of the continuum metric tensor with a simple geometric interpretation. The latter is based on
a correspondence between first order differential calculi and digraphs (the vertices of the latter
are given by the elements of the finite set). Arrows originating from a vertex span its (co)tangent
space. If the metric is to measure length and angles at some point, it has to be taken as an 
element of the {\em left-linear} tensor product of the space of 1-forms with itself, and not as 
an element of the (non-local) tensor product over the algebra of functions, as considered 
previously by several authors. 
It turns out that linear connections can always be extended to this left tensor product, 
so that metric compatibility can be defined in the same way as in continuum Riemannian 
geometry. In particular, in the case of the universal differential calculus on a finite set, the 
Euclidean geometry of polyhedra is recovered from conditions of metric compatibility and 
vanishing torsion. 

In our rather general framework (which also comprises structures which
are far away from continuum differential geometry), there is in general nothing like a
Ricci tensor or a curvature scalar. Because of the non-locality of tensor products (over 
the algebra of functions) of forms, corresponding components (with respect to some
module basis) turn out to be rather 
non-local objects. But one can make use of the parallel transport associated with a
connection to `localize' such objects and in certain cases there is a distinguished way to 
achieve this. In particular, this leads to covariant components of the curvature 
tensor which allow a contraction to a Ricci tensor. Several examples are worked
out to illustrate the procedure. 
 Furthermore, in the case of a differential calculus associated with a hypercubic lattice 
we propose a new discrete analogue of the (vacuum) Einstein equations.}

\section{Introduction}
In a series of papers \cite{DMH92_qlatt,DMH94_fin,DMH94_ddm,DMHV95,BDMHS96}
we have developed a formalism of differential geometry on finite and
discrete sets with applications in particular to lattice gauge theory 
\cite{DMHS93} and discrete completely integrable models \cite{DMH96_id}. 
\vskip.2cm

The most basic `differential geometric' structure on a discrete set $\M$
is a {\em differential calculus} $(\O(\M),\d)$, where 
$\O(\M) = \bigoplus_{r \geq 0} \O^r(\M)$ is an analogue 
of the algebra of differential forms on a differentiable manifold and the
$\mathbb{C}$-linear map $\d \, : \, \O^r(\M) \rightarrow \O^{r+1}(\M)$
generalizes the exterior derivative. Here $\A := \O^0(\M)$ is the algebra 
of $\mathbb{C}$-valued functions on $\M$ and noncommutativity enters the
stage via nontrivial commutation relations between functions and differentials
(which are elements of $\O^1(\M)$). On a discrete set there are many choices
of a (first order) differential calculus and it turned out 
\cite{DMH94_ddm} that these amount to the selection of a digraph structure 
and thus neighbourhood relations on the discrete set. 
\vskip.2cm

Whereas the concept of a connection seems to be well understood in the 
framework of noncommutative geometry, this is not quite so for the concept 
of a metric. In Connes' approach to noncommutative geometry \cite{Conn94}, 
Riemannian geometry is encoded in a selfadjoint operator on a Hilbert space
and recovered from it via a formula for the distance of two points. The
distance formula is then generalized to a more abstract setting, including
the case of discrete sets (see also \cite{DMH97_dist} and references 
therein). A major problem with this approach is that it is bound to
(generalizations of) positive definite metrics and thus at least not
directly applicable to space-time geometry. The underlying
philosophy of `spectral geometry', namely that all geometrical data
should be encoded in the spectrum of certain selfadjoint operators on
a Hilbert space, is certainly very interesting but by no means compulsive.
\vskip.2cm

In several papers (see \cite{BDMHS96,Sita94,DMMM95,Heck+Schm95}, 
for example) a metric in noncommutative geometry has been taken to be an 
element of the tensor product space $\O^1(\A) \otimes_\A \O^1(\A)$ 
with certain properties. Here $\O^1(\A)$ is 
the space of 1-forms of a differential calculus over an associative algebra $\A$.
This has just been a formal generalization of one of several, in classical
differential geometry equivalent, definitions of a metric tensor field,
motivated by simplicity of mathematical structure but without a deeper,
e.g. physical, substantiation. Even on the technical level a serious 
problem showed up, namely the extensibility of a (linear) connection on 
$\O^1(\A)$ to a connection on $\O^1(\A) \otimes_\A \O^1(\A)$, which is 
necessary in order to define metric compatibility of a linear connection
(see \cite{BDMHS96,Dima96} for discussions and related references). 
\vskip.2cm

Needless to say, generalizing another -- classically equivalent -- metric 
concept, one does not in general arrive at equivalent structures in the
noncommutative geometric setting. In fact, motivated by previous work
\cite{DMHS93,DMH96_id} we recently investigated in more detail 
generalizations of the Hodge $\star$-operator \cite{DMH96_deform}.
The metric is recovered from $(\alpha, \beta) = \star^{-1} (\alpha \star
\beta)$ where $\alpha, \beta$ are differential 1-forms. For a symmetric
Hodge operator on a ({\em non}commutative) differential calculus
over a {\em commutative} algebra $\A$, contact was made with a metric
defined as an element 
\be   \label{metric_L}
      g \in \O^1(\A) \otimes_L \O^1(\A)
\ee
and not as an element of the space $\O^1(\A) \otimes_\A \O^1(\A)$. 
The tensor product $\otimes_L$ satisfies
\be         \label{otimes_L}
    (f \, \alpha) \otimes_L (h \, \beta) 
  = f \, h \, (\alpha \otimes_L \beta) \qquad \forall f,h \in \A, \, 
    \alpha, \beta \in \O(\A)  \; . 
\ee
In the following we show that it is precisely the latter metric 
definition which directly reproduces some familiar results in {\em discrete} 
geometry and which allows us to develop discrete noncommutative 
geometry to a more satisfactory level. It should be noticed, however, 
that the tensor product $\otimes_L$ and therefore the metric definition
(\ref{metric_L}) does not generalize in an obvious way to 
{\em non}commutative algebras $\A$, at least as far as we can see. 
But in \cite{DMH96_deform} we have generalized the associated Hodge 
operator to the general noncommutative framework.
\vskip.2cm

In section 2 we recall some basic definitions of noncommutative
geometry.
Section 3 concentrates on finite sets and introduces metrics and compatible 
linear connections on them. 
Section 4 deals with a technical problem which has its origin in the 
non-locality of the tensor product over $\A$. In particular, the construction 
of a Ricci tensor is addressed in our framework.
As an example of particular interest, the 
geometry of a hypercubic lattice is treated in section 5. Section 6
deals with discrete surfaces of revolution.
Some conclusions are collected in section 7. In particular, we propose
a new discrete version of the Einstein equations on a hypercubic lattice.

\section{Preliminaries}
\setcounter{equation}{0}
In the first subsection we recall the definition of a differential
calculus over an associative algebra. The second subsection contains the 
general definitions of linear connections, torsion and curvature in the 
framework of noncommutative geometry. 

\subsection{Differential calculi on associative algebras}
Let $\A$ be an associative algebra over $\mathbb{C}$ 
with unit $\idty$. 
A {\em differential calculus} over $\A$ is a $\mathbb{Z}$-graded 
associative algebra (over $\mathbb{C}$)
\be
           \O (\A) = \bigoplus_{r \geq 0} \O^r (\A)
\ee 
where the spaces $\O^r (\A)$ are $\A$-bimodules and $\O^0(\A) = \A$. 
There is a $\mathbb{C}$-linear map  
\be
     \d \; : \quad \O^r (\A) \rightarrow \O^{r+1}(\A)
\ee
with the following properties,
\be
       \d^2 &=& 0 \\
       \d (w \, w') &=& (\d w) \, w' + (-1)^r \, w \, \d w'
                           \label{Leibniz}
\ee
where $w \in \O^r(\A)$ and $w' \in \O (\A)$. The last relation is
known as the (generalized) {\em Leibniz rule}. One also requires 
$\idty \, w = w \, \idty = w$ for all elements $w \in \O (\A)$. The 
identity $\idty \idty = \idty$ then implies
\be
            \d \idty = 0   \; .
\ee 
 Furthermore, we require that $\d$ generates the spaces $\O^r(\A)$ 
for $r>0$ in the sense that $\O^r(\A) = \A \, \d \O^{r-1}(\A) \, \A$.

\subsection{Linear connections, torsion, and curvature}
Let $(\O(\A),\d)$ be a differential calculus over an associative
algebra $\A$. A linear (left $\A$-module) connection is a 
$\mathbb{C}$-linear map $\nabla \, : \, \O^1(\A) \rightarrow 
\O^1(\A) \oa \O(\A)$ such that
\be           \label{lin_conn}
  \nabla (f \, \alpha) = \d f \oa \alpha + f \, \nabla \alpha 
        \; .
\ee
A linear connection extends to a map $\nabla \, : \, \O(\A) \oa \O^1(\A)
\rightarrow \O(\A) \oa \O^1(\A)$ via
\be 
   \nabla (w \oa \alpha) = \d w \oa \alpha + (-1)^r \, w \, \nabla \alpha  
   \qquad \quad   \forall w \in \O^r(\A) , \, \alpha \in \O^1(\A)   \; .
\ee
\vskip.2cm

The {\em torsion} of a linear connection $\nabla$ is the map
$\Theta \, : \, \O^1(\A) \rightarrow \O^2(\A)$ given by
\be                \label{torsion}
    \Theta(\alpha) := \d \alpha - \pi \circ \nabla \alpha
\ee
where $\pi$ is the natural projection $\O^1(\A) \oa \O^1(\A)
\rightarrow \O^2(\A)$. It satisfies
\be               
    \Theta(f \, \alpha) = f  \, \Theta( \alpha) \; .
\ee
The torsion extends to a map 
$\Theta \, : \, \O(\A) \oa \O^1(\A) \rightarrow \O(\A)$ via
\be
  \Theta(w \oa \alpha) := \d (w \, \alpha) 
          - \pi \circ \nabla(w \oa \alpha)  \qquad
  \forall w \in \O(\A) , \, \alpha \in \O^1(\A)  
\ee
where $\pi$ now denotes more generally the projection 
$\O(\A) \oa \O^1(\A) \rightarrow \O(\A)$. Then
\be
  \Theta(\nabla \alpha) &=& \d \pi \circ \nabla(\alpha) 
      - \pi \circ \nabla^2(\alpha)  \nonumber  \\
  &=& \d (\d \alpha - \Theta(\alpha)) 
           + \pi \circ \mathfrak{R}(\alpha)
\ee
where we have introduced the {\em curvature} $ \mathfrak{R}$ of  
$\nabla$ as the map 
\be
       \mathfrak{R} := - \nabla^2      \label{curv}
\ee
which satisfies 
\be               
    \mathfrak{R}(f \, \alpha) = f \, \mathfrak{R}(\alpha)  \; .
\ee
We arrive at the {\em first Bianchi identity}
\be           \label{1Bianchi}
   \d \circ \Theta + \Theta \circ \nabla = \pi \circ \mathfrak{R}  \; .
\ee
The {\em second Bianchi identity} is
\be
    (\nabla  \mathfrak{R})(\alpha) 
 : = \nabla ( \mathfrak{R}(\alpha)) -  \mathfrak{R} (\nabla \alpha)
   = - \nabla^3 \alpha + \nabla^3 \alpha 
   = 0  \; .
\ee 
\vskip.2cm
\noindent
{\em Example.} For the universal differential calculus, we have 
$\pi = \mbox{id}$ on $\O^1 \oa \O^1$ and there is a unique linear connection
with vanishing torsion given by $\nabla = \d$ according to (\ref{torsion}). 
The curvature of this linear connection vanishes.
                                                                       \hfill $\blacksquare$

\section{Differential geometry on finite sets}
\setcounter{equation}{0}
In this section we collect some facts about differential calculi, 
vector fields and linear connections on finite sets (see also
\cite{DMH94_fin,DMH94_ddm,DMHV95,BDMHS96,Cho+Park97,Zapa97}).
We then consider metrics and elaborate the metric compatibility 
condition for a linear connection.

\subsection{First order differential calculi on a finite set}
Let $\M$ be a finite set of $N$ elements and $\A$ the algebra of all 
$\mathbb{C}$-valued functions on it. $\A$ is a complex linear space
with basis $e^i, \, i=1, \ldots ,N$, where 
$e^i(j) = \delta^i_j$ for $i,j \in \cal M$. These functions satisfy 
the two identities
\be        \label{fin_set_rels}
     e^i \, e^j = \delta^{ij} \,  e^j \, , \qquad 
     \sum_i e^i = \idty  
\ee
where $\idty$ is the constant function on $\M$ with value 1.
In \cite{DMH94_ddm} it has been shown that first order differential
calculi on a finite set $\M$ are in bijective correspondence with
digraph structures on $\M$.
Given a digraph with set of vertices $\M$, we associate with an arrow 
from some point $i$ to another point $j$, denoted as $i \longrightarrow j$
in the following, an algebraic object $e^{ij}$
and define\footnote{Instead of $\O(\A)$ we simply write $\O$ in
the following.}
\be
  \O^1 := \mbox{span}_\mathbb{C} 
  \lbrace e^{ij} \, \mid \, i \longrightarrow j \rbrace \; .
\ee
This is turned into an $\A$-bimodule via
\be          \label{e_e_rels}
    e^i \, e^{kl} = \delta^{ik} \, e^{kl} \, , \quad
    e^{kl} \, e^i = \delta^{li} \, e^{kl}  \; .
\ee
Let us introduce
\be
    \rho = \sum_{k,l} e^{kl}   \label{rho}
\ee
where the summation has to be restricted to those $k,l$ for which there 
is an arrow from $k$ to $l$ in the digraph. Then
\be       \label{df_rho_f}
    \d f = \lbrack \rho , f \rbrack  \qquad  \qquad   f \in \A
\ee
defines a $\mathbb{C}$-linear map $\d \, : \, \A \rightarrow \O^1$
which satisfies the Leibniz rule.  If there is an arrow from $i$ to $j$ in 
the digraph, then $e^i \rho \, e^j = e^{ij}$, otherwise $e^i \rho \, e^j = 0$.
\vskip.2cm

The subspace
\be
      \O^1_i := e^i \, \O^1 
\ee
is generated by the 1-forms $e^{ij}$ corresponding to the arrows originating 
from $i$ in the digraph. It may be regarded as the cotangent space 
at $i \in \M$. We have
\be
      \O^1 = \bigoplus_{i \in \M} \O^1_i  \; .
\ee
\vskip.2cm

The complete digraph where all pairs of points in $\M$ are connected
by a pair of antiparallel arrows corresponds to the largest
first order differential calculus on $\M$, also known as the {\em
universal} first order differential calculus since each other calculus
can be obtained from it as a quotient with respect to some sub-bimodule. 
\vskip.2cm

There is a canonical commutative product in $\O^1$ which satisfies
\be            
     \alpha \bullet \d f = \lbrack \alpha , f \rbrack 
\ee
and 
\be                   \label{f_lin_bullet}
 (f \, \alpha \, f') \bullet ( h \, \beta \, h') = f  h \, (\alpha \bullet \beta) \, f' h' 
 \qquad \qquad 
 \forall f, f', h, h' \in \A, \, \alpha, \beta \in \O^1  \; .
\ee
More generally, this product exists for every first order differential
calculus over a commutative algebra \cite{BDMH95}. In the case
under consideration, it is given by
\be            \label{bullet}
  e^{ij} \bullet e^{kl} = \delta^{ik} \, \delta^{jl} \, e^{ij}  \; .
\ee

The space of 1-forms $\O^1$ is free as a (left or right) $\A$-modul.
A special left $\A$-module basis is given by 
\be    
     \rho^i := \sum_j \, e^{ji}    \qquad   \mbox{if }  \rho \, e^i \neq 0
\ee
since an arbitrary 1-form $A$ can be written as
\be
     A = \sum_{ij} A_{ij} e^{ij} = \sum_i A_i \, \rho^i
\ee
where $A_i = \sum_j A_{ji} \, e^j$. Furthermore, $\sum_i A_i \, \rho^i = 0$ implies, 
via multiplication with $e^j$ from the left, that $A_{ji}=0$ and thus $A_i =0$.

\subsection{Higher order differential forms on a finite set}
Concatenation of the 1-forms $e^{ij}$ leads to the $r$-forms
\be                         \label{e_...}
  e^{i_0 \ldots i_r} := e^{i_0 i_1} \, e^{i_2 i_3} \cdots e^{i_{r-1} i_r}  
        \qquad  ( r > 0 ) 
\ee 
which can also be expressed as follows, 
\be              \label{e_rho_e}
   e^{i_0 \ldots i_r} = e^{i_0} \, \rho \, e^{i_1} \, \rho
   \, \cdots \, \rho \, e^{i_r} \; .
\ee
They satisfy the simple relations
\be                          \label{e_e_delta_e}
       e^{i_0 \ldots i_r} \, e^{j_0 \ldots j_s} 
   = \delta^{i_r j_0} \, e^{i_0 \ldots i_{r-1} j_0 \ldots j_s}
\ee
and span $\O^r$ as a vector space over $\mathbb{C}$. Using 
(\ref{e_e_rels}) this space is turned into an $\A$-bimodule. The
exterior derivative $\d$ extends to higher orders via 
\be      
  \d e^i &=& \rho \, e^i - e^i \, \rho  \label{de^i}   \\
  \d \rho &=& \rho^2 + \sum_i e^i \, \rho^2 \, e^i    \label{drho}
\ee
and the (graded) Leibniz rule (\ref{Leibniz}). In particular, this leads to
\be      
  \d e^{ij} &=& \rho \, e^i \, \rho \, e^j - e^i \, \rho^2 e^j + e^i \, \rho \, e^j \, \rho  
                            \label{de^ij}   \\
  \d e^{ijk} &=& \rho \, e^i \, \rho \, e^j \, \rho \, e^k 
                          - e^i \, \rho^2 e^j \, \rho \, e^k 
                          + e^i \, \rho \, e^j \, \rho^2 e^k
                          - e^i \, \rho \, e^j \, \rho \, e^k \, \rho   \; .   \label{de^ijk}
\ee
\vskip.2cm

Starting with the universal first order differential calculus on $\M$,
these formulas generate the {\em universal differential calculus}
(which is also known as the 
{\em universal differential envelope} of $\A$). A smaller first order
differential calculus (where some of the $e^{ij}$ are missing)
induces restrictions on the spaces of higher order forms. 
A missing arrow from $i$ to some other point $j$ (in the complete digraph 
on $\M$) means $e^i \rho \, e^j = 0$. Acting with $\d$ on this equation, using 
(\ref{de^i}) and (\ref{drho}), leads to
\be
   i \, \not \! \! \longrightarrow \, j
   \quad \Rightarrow \quad
   e^i \, \rho^2 \, e^j = 0  \; .
\ee
Each differential calculus is obtained from the universal one as a quotient with respect 
to some differential ideal. If the differential ideal is generated by `basic forms' (\ref{e_...})
only\footnote{In general, a differential ideal is generated by linear combinations of basic 
forms.}, then the differential calculus is called {\em basic} \cite{Zapa97}. This class
of differential calculi has been associated with polyhedral representations of simplicial
complexes \cite{Zapa97}.

\subsection{Vector fields on a finite set}
Let $\X$ denote the dual of $\O^1$ as a complex vector space. 
Let $\lbrace \pa_{ji}\rbrace$ be the basis of $\X$ dual to 
$\lbrace e^{ij} \rbrace$. If $\langle \, , \rangle_0$ denotes
the duality contraction, then 
\be  
  \langle e^{ij}, \pa_{kl}\rangle_0 = \delta^i_l \, \delta^j_k  \; .
\ee
$\X$ is turned into an $\A$-bimodule by introducing the left and 
right actions
\be
  \langle \alpha, f \cdot X \rangle_0 := \langle \alpha f, X \rangle_0 \; ,
  \qquad
  \langle \alpha, X \cdot f \rangle_0 := \langle f \alpha, X \rangle_0  \; .
\ee
As a consequence,
\be 
  e^k \cdot \pa_{ji}  =\delta^k_j \, \pa_{ji} \, , \qquad
   \pa_{ji} \cdot e^k = \delta^k_i \, \pa_{ji}  \; .
\ee
An element $X \in \X$ can be uniquely decomposed as follows,
\be
     X = \sum_{i \longrightarrow j} X(i)^j \, \pa_{ji}
\ee
(where the summation runs over all $i,j \in \M$ for which there is an arrow
from $i$ to $j$ in the digraph associated with $\O^1$). 
Now we introduce a duality contraction 
$\langle \, ,  \rangle$ of $\O^1$ as a right $\A$-module and $\X$ as a left
$\A$-module by setting
\be
     \langle e^{ij}, X \rangle := e^i \, \langle e^{ij}, X \rangle_0  
\ee
for all $X \in \X$. Then we have
\be
 \langle f \alpha , X \cdot h \rangle = f \, \langle \alpha , X \rangle \, h \, ,
 \qquad
 \langle \alpha , f \cdot X \rangle = \langle \alpha f , X \rangle \; .
\ee
The elements of $\X$ become operators on $\A$ via
\be
   X(f) := \langle \d f , X \rangle  \; .
\ee
Using the Leibniz rule for $\d$, one proves
\be
    X(fh) = f \, X(h) + (h \cdot X)(f)  \qquad  \qquad \forall f,h \in \A \; .
\ee
 Furthermore, 
\be
    (X \cdot f)(g) =  X(g) \, f   \; .
\ee 
The duality contraction extends to the pair of spaces 
$\O \oa \O^1$ and $\X \oa \O$ via
\be
  \langle w \oa \alpha , X \oa w' \rangle =
  w \, \langle \alpha , X \rangle \, w'  \; .
\ee
\vskip.2cm

The space
\be
     \X_i := \X \, e^i = \lbrace X \cdot e^i \, | \, X \in \X \rbrace 
\ee
may be regarded as the tangent space at $i \in \M$. 
It is dual to $\O_i^1$ with respect to the duality contraction 
$\langle \; , \, \rangle_0$.
The set $\lbrace \pa_{ji}\, | \, j \in \M \; \mbox{such that}\; 
i \longrightarrow j \rbrace$ is
a basis of $\X_i$ which is dual to the basis  $\lbrace e^{ij}\, | \, j \in \M \; 
\mbox{such that}\; i \longrightarrow j \rbrace$ of $\O_i^1$.

\subsection{Linear connections on a finite set}
Let $\nabla \, : \, \O^1 \rightarrow \O^1 \oa \O^1$ be a (left $\A$-module) 
linear connection. Using (\ref{lin_conn}) and the properties of $\rho$, 
one finds that
\be
  \U(\alpha) := \rho \oa \alpha - \nabla \alpha     \label{U_rho_nabla}
\ee
is a left $\A$-homomorphism $\U \, : \, \O^1 \rightarrow \O^1 \oa \O^1$, i.e.,
\be                     \label{U_left_linear}
   \U(f \alpha) = f \, \U(\alpha)  \qquad  \forall f \in \A, \, \alpha \in \O^1  \; .                                              
\ee
We call $\U$ the {\em parallel transport} associated with the linear 
connection $\nabla$. 
In particular, (\ref{U_left_linear}) implies $\U(e^{ij}) = e^i \, \U(e^{ij})$ 
and thus we have an expansion
\be              \label{U_components}
   \U(e^{ij}) = \sum_{k,l} U(i)^j{}_{kl} \, e^{ik} \oa e^{kl}
 = \sum_k e^{ik} \oa \sum_l U(i)^j{}_{kl} \, e^{kl}
\ee
with constants $U(i)^j{}_{kl}$.
\vskip.2cm

Via
\be                \label{U^ij}
      e^{ik} \quad \mapsto \quad (e^{ik}) \U^{ij} := \sum_l U(i)^k{}_{j l} \, e^{j l}
\ee
for fixed $i$ and $j$, the parallel transport defines a linear map $\O^1_i \rightarrow \O^1_j$ 
with associated matrix $\U^{ij}$. Then we have
\be
      \U(\alpha) = \sum_{i,j}e^{ij}\oa [(e^i \, \alpha) \U^{ij}]  \; .
\ee
\vskip.2cm

Given a linear connection on $\O^1$, there is a dual connection\footnote{We
use the same symbol $\nabla$ for the connection and its dual.} 
$\nabla \, : \, \X \rightarrow \X \oa \O^1$, such that
\be
  \d \langle \alpha , X \rangle = \langle \nabla \alpha , X \rangle +
  \langle \alpha , \nabla X \rangle  
\ee
(cf \cite{BDMHS96}, appendix B).
Using $\d \langle \alpha ,X \rangle = [ \rho , \langle \alpha , X \rangle]$ 
one proves that the dual parallel transport defined by
\be
     \langle \alpha , \U(X) \rangle = \langle \U(\alpha) , X \rangle
\ee
acts as follows on $\X$,
\be
     \U(X) := X \oa \rho + \nabla X  \, ,
\ee
and satisfies 
\be
   \U(X \cdot f) = \U(X) \, f  \; .
\ee
(\ref{U_components}) leads to
\be           \label{U_of_pa}
     \U(\pa_{ji}) = \sum_{k,l} U(k)^l{}_{ij} \, \pa_{l k}\oa e^{ki}  \; .
\ee
The parallel transport (and thus also the connection) extends in an obvious way to 
$\O \oa \O^1$ and $\X \oa \O$ as graded left respectively right $\O$-homomorphisms,
i.e.,
\be
     \U(w \oa \alpha) = (-1)^r \, w \oa \U(\alpha) \; , \qquad
     \U(X \oa w) = (-1)^r \, \U(X) \oa w 
\ee
where $w \in \O^r$.
\vskip.2cm

The map $ \X_j \rightarrow \X_i$ dual to the parallel transport map with
matrix $\U^{ij}$ defined in (\ref{U^ij}) is given by
\be
     \pa_{ki} \quad \mapsto \quad   \sum_l U(j)^l_{ik} \, \pa_{lj} 
        = \U^{ij}(\pa_{ki})     \; .
\ee
Now (\ref{U_of_pa}) extends to
\be          
     \U(X) = \sum_{i,j} \U^{ij}(X \cdot e^i) \oa e^{ij}  \; .
\ee
\vskip.2cm

We may introduce the curvature as the right $\O$-homomorphism
$\R' \, : \, \X \oa \O \rightarrow \X \oa \O$ defined by	 
\be 
         \R' = \nabla^2  \; .
\ee
Its dual $\R : \O \oa \O^1 \rightarrow \O \oa \O^1$ is then given by $\R= - \nabla^2$ 
in accordance with our general definition (\ref{curv}). We obtain
\be
   \R(e^{ij}) &=:& \sum_{k,l,m} R(i)^j{}_{klm} \, 
                      e^{ikl} \oa e^{lm}   \nonumber \\
    &=& \sum_{k,l,m}\left( \sum_n U(i)^j{}_{kn}\,U(k)^n{}_{l m} 
                      - U(i)^j{}_{l m}\right) \, e^{ikl}\oa e^{l m}
                        \label{curv_comp}
\ee
where it has been convenient to set 
\be
       U(i)^j{}_{ik} := \delta^j_k  \; .   
\ee
We also have the following expression for the curvature,
\be
   \R(\alpha) = \sum_{i,j,k}e^{ijk}\oa \bigl \lbrace (e^i \, \alpha) 
      [\U^{ij} \, \U^{jk} - \U^{ik}] \bigr \rbrace
\ee
where $\U^{ii} :={\rm id}_{\O^1_i}$. 
\vskip.2cm

 For the torsion we find
\be
      \Theta(e^{ij}) 
  = - e^i \rho^2 e^j + e^{ij}\, \rho + \sum_{k,l} U(i)^j{}_{kl}\, e^{ikl}
  = \sum_{k,l}( \delta^j_k - \delta^j_l + U(i)^j{}_{kl}) \, e^{ikl}  \; .
\ee

\vskip.2cm
\noindent
{\em Example.} In case of the universal differential calculus, the condition 
of vanishing torsion leads to
\be              \label{univ_lin_conn}
       U(i)^j{}_{kl} = \delta^j_l - \delta^j_k
\ee
and thus fixes the linear connection completely.\footnote{This is no
longer so when $\O^2$ is smaller than $\O^1 \oa \O^1$.} 
As mentioned in more generality in the example in section 2.2, this
connection is given by $\nabla = \d$ and its curvature vanishes.
                                       \hfill $\blacksquare$

\subsection{Metrics and compatible linear connections on finite sets}
Using
\be
    e^{ij} \otimes_L e^{kl} = e^{ij} \otimes_L e^k \, e^{kl}
 = e^k \, e^{ij} \otimes_L e^{kl} = \delta^{ki} \, e^{ij} \otimes_L e^{il}
\ee
one finds that an element $g \in \O^1 \otimes_L \O^1$ can be 
expressed as
\be       \label{g_leftcov}
     g = \sum_{i,j,k} g(i)_{jk} \, e^{ij} \otimes_L e^{ik} 
\ee
with constants $g(i)_{jk}$. This will be our candidate for a 
{\em metric} on $\M$.\footnote{At this point it is worth not to 
impose additional conditions. Finally we will be interested in $g$ being
{\em real} and {\em symmetric} (i.e., $g(i)_{jk} = g(i)_{kj}$), or {\em Hermitean}.
We refer to $g(i)_{jk}$ as the components of a `metric' at $i$ in order to emphasize
a certain analogy with a metric tensor in continuum differential geometry. However,
a better name would be {\em distance matrix} of the digraph at $i$. In general,
$g(i)$ will be degenerate.}

\vskip.2cm
\noindent
{\em Example 1.} Consider a digraph embedded in Euclidean space
such that the arrows are straight lines of Euclidean length 
$\ell_{ij}$. Let $\vartheta_{jik}$ denote the angle between
arrows from $i$ to $j$ and from $i$ to $k$. Define\footnote{More generally,
let us consider a graph embedded in some affine space ${\mathbb R}^d$, 
$d \in {\mathbb N}$, with inner product $(\; , \,)$. Hence, there is a map 
$\vec{x} \; : \; \M \rightarrow \mathbb{R}^d$ with 
$\vec{x} = \sum_{i \in \M} \vec{x}_i \, e^i$. Given a (first order) differential 
calculus on $\M$, we have 
$\d \vec{x} = \sum_{i,j} (\vec{x}_j - \vec{x}_i) \, e^{ij}$. 
The inner product then induces a metric on $\M$ via 
$ g(i)_{jk} = (\vec{x}_j - \vec{x}_i , \vec{x}_k - \vec{x}_i)$.
If the inner product is the Euclidean one, then we have (\ref{Euclid_g}).}
\be            \label{Euclid_g}
   g(i)_{jj} = \ell_{ij}^2 \, , \qquad 
   g(i)_{jk} = \ell_{ij} \, \ell_{ik} \, \cos \vartheta_{jik}
               \; .
\ee
In order to describe the geometry of a polygon (without orientation
of its lines) embedded in Euclidean space completely, in general we need 
to associate it with a {\em symmetric} digraph. A line between two
points $i$ and $j$ is then represented by a pair of antiparallel
arrows, so that $e^{ij}$ and $e^{ji}$ are both present.
Of course, we should impose $\ell_{ij} = \ell_{ji}$.\footnote{Our 
formalism admits non-standard geometries, however. For example,
measuring the (not necessarily spatial) `distances' from $i$ to $j$ and from $j$ to $i$
in some (in a generalized sense) anisotropic space may lead to different results. 
This can be taken into account by dropping the restriction $\ell_{ij} =
\ell_{ji}$.}
                                   \hfill $\blacksquare$
\vskip.2cm

In order to define compatibility of a linear connection and 
a metric, we have to extend the connection, respectively the
map $\U$, from $\O^1$ to $\O^1 \otimes_L \O^1$. Let us 
define
\be
   \U(\alpha \otimes_L \beta) := \bullet \left( \U(\alpha) \otimes_L 
   \U(\beta) \right)
\ee
where a map 
\be
 \bullet \, : \, (\O^1 \oa \O^1) \otimes_L (\O^1 
 \oa \O^1) \rightarrow \O^1 \oa (\O^1 \otimes_L  \O^1)    
\ee
is needed. Using the canonical product (\ref{bullet}) in the space
of 1-forms, such a map is given by
\be
   \bullet \left( (\alpha \oa \beta) \otimes_L (\alpha' \oa
                  \beta') \right)
    := (\alpha \bullet \alpha') \oa (\beta \otimes_L \beta')
\ee
and, using (\ref{f_lin_bullet}), we have
\be
   \U(f \, (\alpha \otimes_L \beta) ) =
   f \, \U(\alpha \otimes_L \beta)  \; .
\ee
As a consequence, 
\be
 \nabla(\alpha \otimes_L \beta) := \rho \oa (\alpha \otimes_L
 \beta) - \U(\alpha \otimes_L \beta)
\ee
defines a (left $\A$-module) connection on $\O^1 \otimes_L \O^1$. 
The {\em metric compatibility} condition $\nabla g = 0$ now
amounts to
\be            \label{rho_U_metric-comp}
         \rho \oa g = \U(g)  \; .
\ee
In terms of the matrices $\U^{ij}$ introduced in section 3.4, we have
\be 
    \U(\alpha \otimes_L \beta) = \sum_{i,j}e^{ij}\oa \bigl \{ 
  [(e^i \alpha) \U^{ij}] \otimes_L (e^i \, \beta) \U^{ij} ] \bigr \}  \; .
\ee

\vskip.2cm
\noindent
{\em Lemma.} Expressed in components, $\nabla g = 0$ becomes
\be         \label{g_compat_components}
  g(i)_{jk} = \sum_{m,n} g(l)_{mn} \, U(l)^m{}_{ij} \,
             U(l)^n{}_{ik} 
\ee
for all $i,l \in \M$ such that $l \longrightarrow i$ (i.e., there is 
an arrow from $l$ to $i$ in the digraph associated with $\O^1$).  
\vskip.1cm \noindent
{\em Proof.} 
\bea
  \U(g) &=& \sum_{l,m,n} g(l)_{mn} \, \bullet( 
       \U(e^{lm}) \otimes_L \U(e^{ln}) )  \\
 &=& \sum_{l,m,n} g(l)_{mn} \, \sum_{i,j,k,p} U(l)^m{}_{ij}
     \, U(l)^n{}_{pk} \, \bullet((e^{li} \oa e^{ij}) \otimes_L
     (e^{lp} \oa e^{pk}))  \; .
\eea
With
\bea
  \bullet((e^{li} \oa e^{ij}) \otimes_L (e^{lp} \oa e^{pk}))
 &=& (e^{li} \bullet e^{lp}) \oa (e^{ij} \otimes_L e^{pk})  \\
 &=& \delta^{ip} \, e^{li} \oa (e^{ij} \otimes_L e^{pk})
\eea
this becomes
\bea
 \U(g) = \sum_{i,j,k,l,m,n} g(l)_{mn} \, U(l)^m{}_{ij}
 \, U(l)^n{}_{ik} \, e^{li} \oa (e^{ij} \otimes_L e^{ik})  \; .
\eea
Using (\ref{rho_U_metric-comp}), the last expression must be equal to
\bea
   \rho \oa g = \sum_{i,j,k,l} g(i)_{jk} \, e^{li} \oa
                (e^{ij} \otimes_L e^{ik})  \; .
\eea
Comparison of the coefficients on both sides now leads to our formula.
                   \hfill $\blacksquare$

\vskip.2cm
\noindent
{\em Example 2.} Again, we consider the universal differential
calculus on $\M$. With the unique torsion-free linear connection
(\ref{univ_lin_conn}), the metric compatibility condition 
reads\footnote{Note that $g(i)_{ik}$ and $g(i)_{ki}$ do not appear in 
(\ref{g_leftcov}) and have to be interpreted as $0$ in the following formulas.}
\be   \label{uc_metric_comp}
  g(i)_{kl} = g(j)_{kl} + g(j)_{ii} - g(j)_{ki} - g(j)_{il} 
  \qquad  i,j,k,l \in \M  \; .
\ee
Setting $k=j$ and $l=j$, respectively, we get
\be           \label{uc_mc_1}
    g(i)_{jk} = g(j)_{ii} - g(j)_{ik} \, , \qquad
    g(i)_{kj} = g(j)_{ii} - g(j)_{ki}
\ee
which in turn implies
\be            \label{uc_mc_2}
    g(i)_{jk} - g(i)_{kj} = g(j)_{ki} - g(j)_{ik} 
\ee
and
\be           \label{uc_mc_3}
    g(i)_{jj} = g(j)_{ii} \; .
\ee
 Furthermore, the last equation together with (\ref{uc_metric_comp})
leads to
\be
    2 \, g(i)_{kl} - g(i)_{kj} - g(i)_{jl}
  = 2 \, g(j)_{kl} - g(j)_{ki} - g(j)_{il}
\ee
which for $k=l$ becomes
\be           \label{uc_mc_4}
    2 \, g(i)_{kk} - g(i)_{kj} - g(i)_{jk}
  = 2 \, g(j)_{kk} - g(j)_{ki} - g(j)_{ik}  \; .
\ee
Let us now consider the special case where all the components
$g(i)_{jj}$ are equal. Then (\ref{uc_mc_2}) and (\ref{uc_mc_4})
lead to
\be
     g(i)_{kj} = g(j)_{ik}  \; .
\ee
With the help of (\ref{uc_mc_1}) and (\ref{uc_mc_3}) we now
obtain
\be
    g(i)_{jj} = g(i)_{jk} + g(i)_{kj}  \; .
\ee
Assuming in addition that the metric is symmetric (i.e., 
$g(i)_{jk} = g(i)_{kj}$), we have
\be
    g(i)_{jj} = 2 \, g(i)_{jk}
\ee
and we end up with a constant metric 
\be
    g(i) = \left( \begin{array}{cccc}
                  a      & a/2    & \ldots & a/2    \\
                  a/2    & \ddots &        & \vdots \\
                  \vdots &        & \ddots & \vdots \\ 
                  a/2    & \cdots & a/2    & a
                  \end{array}  \right)
    \qquad \qquad \forall i \in \M  \; .    \label{reg_pol_g}
\ee
Hence, there is a unique symmetric $g$ for the universal differential calculus 
(associated with the complete digraph) on $\M$ which is 
compatible with the (unique) torsion-free linear connection and 
which has the property that all $g(i)_{jj}$ are equal. 
If $g(i)_{jj}$ is positive, we let it represent the square of
the distance between $i$ and $j$. The above requirement then 
means that all points are at equal distance $\ell = \sqrt{a}$
and from the metric compatibility condition we recover the
Euclidean geometry of the regular polyhedron.

More generally, specializing to the `Euclidean metric' 
(\ref{Euclid_g}), our metric compatibility conditions
(\ref{uc_metric_comp}) become
\be
   \ell_{ik}^2 &=& \ell_{jk}^2 + \ell_{ji}^2 
   - 2 \, \ell_{ji} \ell_{jk} \, \cos(\vartheta_{ijk}) \label{Eucl_triangle} \\
   \ell_{ik} \ell_{il} \, \cos(\vartheta_{kil}) &=&
   \ell_{jk} \ell_{jl} \, \cos(\vartheta_{kjl}) + \ell_{ji}^2
   - \ell_{ji} \ell_{jk} \, \cos(\vartheta_{ijk}) 
   - \ell_{ji} \ell_{jl} \, \cos(\vartheta_{ijl})  \quad
\ee
which in fact reproduce well-known relations of Euclidean 
geometry.
                                \hfill $\blacksquare$

\vskip.2cm
\noindent
In terms of the matrices
\be
                    g(i) := (g(i)_{jk})
\ee
the metric compatibility condition takes the simple form
\be
            g(j) = (\U^{ij})^t \, g(i) \, \U^{ij} 
\ee
where $(\U^{ij})^t$ denotes the transpose of the matrix $\U^{ij}$. 
Hence, if there is an arrow from $i$ to $j$ in the digraph (i.e., 
$i \longrightarrow j$), then $g(i)$ determines $g(j)$ via the parallel transport 
of a metric compatible linear connection.
\vskip.2cm

The metric compatibility condition implies that, for any closed path 
$i_0 \longrightarrow i_1 \longrightarrow \ldots \longrightarrow i_r 
\longrightarrow i_0$ in the digraph, the matrix 
$\H^{i_0 \ldots i_r} := \U^{i_0 i_1} \U^{i_1 i_2} \cdots \U^{i_r i_0}$ 
must be in the orthogonal group of $g(i_0)$. The set of all matrices
$\H^{i_0 \ldots i_r}$, $r \geq 1$, forms the holonomy group $G_H(i_0)$
at $i_0 \in \M$.

\vskip.2cm
\noindent
{\em Example 3.} The three point complete digraph. \\
Let $\M = \{1,2,3\}$ with 
$\rho = e_{12}+e_{13}+e_{21}+e_{23}+e_{31}+e_{32}$. 
We are dealing again with the universal differential calculus
so that there are no 2-form relations. Then
$\rho^2 = e_{121}+e_{123}+e_{131}+e_{132}+e_{212}+e_{213}+e_{231}
+e_{232}+e_{312}+e_{313}+e_{321}+e_{323}$.
The condition of vanishing torsion determines the connection
completely. We find
\be
 \U^{12} = \left(\begin{array}{cc} -1 & -1 \\ 0 & 1 \end{array}\right) \, ,
 \qquad
 \U^{13} = \left(\begin{array}{cc} 0 & 1 \\ -1 & -1 \end{array}\right) \, ,
 \qquad
 \U^{23} = \left(\begin{array}{cc} 1 & 0 \\ -1 & -1 \end{array}\right) \; \;
                \nonumber \\
 \U^{21} = \left(\begin{array}{cc} -1 & -1 \\ 0 & 1 \end{array}\right)
  \, ,            \qquad
 \U^{31} = \left(\begin{array}{cc} -1 & -1 \\ 1 & 0\end{array}\right) \, ,
 \qquad
 \U^{32} = \left(\begin{array}{cc} 1 & 0 \\ -1 & -1 \end{array}\right) \, .
\ee
It follows that $\H^{ij} = I$, the unit matrix, for all $i \longrightarrow j 
\longrightarrow i$.  Furthermore, for all permutations $i,j,k$ of 1,2,3 we find
$\H^{ijk} = \U^{ij} \U^{jk} \U^{ki} = I$. This means that parallel transport does
not depend on the path which is related to the fact that the curvature vanishes.
If we choose metric components at one point, then the metric components
at the other points are determined via the metric compatibility condition.
We find
\be
 g(1)=\left(\begin{array}{cc} a & b \\ b & c\end{array}\right),   \;
 g(2)=\left(\begin{array}{cc} a & a-b \\ a-b &
 a-2b+c\end{array}\right),    \;
 g(3)=\left(\begin{array}{cc} c & c-b \\ c-b &
 a-2b+c\end{array}\right)  \, .
\ee
In particular, if $g(1)=g(2)=g(3)$ we are led to
\be
 g(i) =
 b \, \left(\begin{array}{cc} 2 & 1 \\ 1 & 2 \end{array} \right)
\ee
(in accordance with (\ref{reg_pol_g}))
which (for $b > 0$) describes an equilateral triangle. This may be considered
as a simple model of a piece of a 2-dimensional surface. 
                                \hfill $\blacksquare$
\vskip.2cm

Thinking about an inverse (or dual) of a metric tensor, as defined above,
one is led to elements $h \in \X \otimes_R \X$ where $\otimes_R$ denotes the 
{\em right} linear tensor product. $h$ can be expressed as
\be
    h = \sum_{i,j,k} h(i)^{jk}\, \pa_{ji}\otimes_R \pa_{ki}
\ee
with constants $h(i)^{jk}$. The parallel transport (and thus also the
connection) extends to $\X \otimes_R \X$ via
\be
     \U(X \otimes_R Y) := \bullet \left(\U(X) \otimes_R \U(Y) \right)
\ee
and
\be
   \bullet \left((X \oa \alpha) \otimes_R (Y \oa \beta) \right) 
  := (X \otimes_R Y) \oa (\alpha \bullet \beta) \; .
\ee
Compatibility of $h$ with a linear connection, i.e., $\nabla h = 0$,  now reads
\be
        \U(h) = h \oa \rho  
\ee
and, in components,
\be
     h(i)^{rs} = \sum_{j,k} h(l)^{jk}\,U(i)^r{}_{lj}\, U(i)^s{}_{lk} 
\ee
provided that $i \longrightarrow l$.
In terms of the matrices $h(i) := (h(i)^{jk})$, the metric compatibility 
condition reads
\be      \label{h_compat}
          h(i) = \U^{ij} \, h(j) \, (\U^{ij})^t  \; .
\ee
\vskip.2cm
\noindent
{\em Remark.} Consider a differential calculus, associated with a
symmetric digraph, a metric $g$ and a compatible linear connection. 
If $g(i_0)$ is invertible at some point $i_0$, setting $h(i_0) := g(i_0)^{-1}$ 
defines $h$ via (\ref{h_compat}) on the connected component of the digraph
containing $i_0$. Of course, $h$ need not be inverse to $g$ at other points.
                                                   \hfill $\blacksquare$

\subsubsection{... with a basic differential calculus}
We consider a {\em basic} differential calculus (cf section 3.2).
The general torsion-free connection is then given by
\be
     U(i)^j{}_{kl} = \delta^j_l-\delta^j_k + u(ikl)^j   \label{Uu}
\ee
where $u(ikl)^j \neq 0$ only if $e^{ikl} = 0$.\footnote{Here 
``if $e^{ikl} = 0$'' should be interpreted as ``if $e^{ikl}$ is not
present in the differential calculus''. This abuse of notation has the
great advantage of being much more concise and will
therefore be repeatedly used in the following.} 
The metric compatibility condition now becomes 
\be
 g(j)_{kl} &=& g(i)_{kl} - g(i)_{kj} - g(i)_{jl} + g(i)_{jj} \nonumber \\
 & & + \sum_{m,n} g(i)_{mn} \, [\delta^m_k \, u(ijl)^n+\delta^n_l \,
     u(ijk)^m + u(ijk)^m \, u(ijl)^n]                        \nonumber \\ 
 & & -\sum_m [g(i)_{jm} \, u(ijl)^m+g(i)_{mj} \,u(ijk)^m] 
                                                \label{mc}
\ee 
for all $i,j$ with $i \longrightarrow j$.
\vskip.2cm
\noindent
{\em Remark.}
Let us consider again the case of a Euclidean embedding space (cf example 1).
If all $u(ijk)^l$ vanish, then (\ref{Eucl_triangle}) holds
which is a familiar relation between the lengths and angles of a Euclidean triangle.
As shown in \cite{Brew97}, in the triangulation of a curved space by means of 
geodesic segments and in Riemann normal coordinates one has
\be
  2 \, \ell_{ij} \ell_{ik} \cos \vartheta_{jik} = \ell_{ik}^2 + \ell_{ij}^2 - \ell_{jk}^2
  - {1 \over 3} \, R_{\mu \alpha \nu \beta} \, \Delta x^\mu_{ij} \,
  \Delta x^\nu_{ij} \, \Delta x^\alpha_{ik} \Delta x^\beta_{ik}
 + {\cal O}(\epsilon^5)
\ee
where $\epsilon$ is a typical length scale of the neighbourhood in
which the Riemann normal coordinates are defined, and $x^\mu_i$ are
the Riemann normal coordinates of the vertex $i$.
Obviously, from (\ref{mc}) we can expect to get additional terms  
in (\ref{Eucl_triangle}), related to curvature, only if we have nonvanishing 
$u(ijk)^\ell$, that is if we have 2-form relations as in our next example. 
                                                           \hfill $\blacksquare$

\vskip.2cm
\noindent
{\em Example 4.} 
A refined model for a piece of a 2-dimensional surface  is obtained from that 
considered in example 3 by adding a fourth point to the triangle and joining it with 
all the vertices of the latter, but then discard the 2-forms 
corresponding to the base of the resulting tetrahedron (or pyramid with triangle base). 
Hence we consider the complete digraph on $\M = \{1,2,3,4\}$, but
not the universal differential calculus since we impose the 2-form
relations
\be
   e^{123}=e^{132}=e^{213}=e^{231}=e^{312}=e^{321}=0 \; .
\ee
We assume that the matrices $\U^{ij}$ have maximal rank and that 
\be  
     \H^{ij} = \U^{ij} \U^{ji} = I  \; .      \label{triv_holon}
\ee
The condition of vanishing torsion now leads to
\be
 \U^{12} \! &=& \!
 \left(\begin{array}{ccc} 
 -1 & -1+u_1 & -1 \\ 0 & 1+u_2 & 0 \\  0 & u_3 & 0
 \end{array}\right) \, ,
 \qquad
 \U^{13} =
 \left(\begin{array}{ccc} 
 0 & 1+v_1 & 0 \\ -1 & -1+v_2 & -1 \\  0 & v_3 & 1
 \end{array}\right) \, ,    \nonumber \\
 \U^{23} \! &=& \!
 \left(\begin{array}{ccc} 
 1+w_1 & 0 & 0 \\ -1+w_2 & -1 & -1 \\  w_3 & 0 & 1
 \end{array}\right) \, ,
 \qquad
 \U^{14} =
 \left(\begin{array}{ccc} 
 0 & 1 & 0 \\ 0 & 0 & 1 \\  -1 & -1 & -1
 \end{array}\right) \, ,  \nonumber \\
 \U^{24} \! &=& \!
 \left(\begin{array}{ccc} 
 1 & 0 & 0 \\ 0 & 0 & 1 \\  -1 & -1 & -1
 \end{array}\right) \, ,
 \qquad \qquad \; \,
 \U^{34} =
 \left(\begin{array}{ccc} 
 1 & 0 & 0 \\ 0 & 1 & 0 \\  -1 & -1 & -1
 \end{array}\right) 
\ee
and for $i<j$ we have $\U^{ji} = (\U^{ij})^{-1}$ according to 
(\ref{triv_holon}). Setting
\be
   g(4) = \ell^2 \left(\begin{array}{ccc} 
1 & c & b \\ c & 1 & a \\ b & a & 1 \end{array}\right) 
\ee
means that the edges of the triangles 4-1-2, 4-1-3, 4-2-3 have equal length
$\ell_{41} = \ell_{42} = \ell_{43} = \ell$ but possibly different angles 
$\cos\th_{142} = c ,\, \cos\th_{143} = b, \, \cos\th_{243}=a$. 
Via $g(i) = (\U^{4i})^t \, g(4) \, \U^{4i}$ for $i=1,2,3$ we obtain
\be
   g(1) &=& \ell^2 \left(\begin{array}{ccc}
  2 \, (1-c) & 1+a-b-c & 1-c \\1+a-b-c & 2 \, (1-b) & 1-b\\ 1-c & 1-b & 1
  \end{array}\right)    \nonumber \\
  g(2) &=& \ell^2 \left(\begin{array}{ccc}
  2 \, (1-c) & 1-a+b-c & 1-c \\ 1-a+b-c & 2 \, (1-a) & 1-a \\ 1-c & 1-a & 1
 \end{array}\right)         \nonumber \\
  g(3) &=& \ell^2 \left(\begin{array}{ccc}
  2 \, (1-b) & 1-a-b+c & 1-b \\ 1-a-b+c & 2 \, (1-a) & 1-a \\ 1-b & 1-a & 1
 \end{array}\right) \; .
\ee
The remaining metric compatibility conditions now demand that 
\be
  u_2=v_1=w_1=-2 \; , \quad
  u_1 = 2 \, {bc-a \over c^2-1} \; , \quad
  v_2 = 2 \, {bc-a \over b^2-1} \; , \quad
  w_2 = 2 \, {ac-b \over a^2-1} 
\ee
and
\be
 u_3 = 2 \, {1-a-b+c \over 1+c} \; , \qquad
 v_3 = 2 \, {1-a+b-c \over 1+b} \; , \qquad
 w_3 = 2 \, {1+a-b-c \over 1+a} 
\ee
where we assumed that $a,b,c \neq \pm1$. We should mention here that
$u_1 = \cdots = w_3 =0$ is also a solution. This parallel transport,
which corresponds to the unique torsion-free connection on the universal 
differential calculus on the set of four points, has vanishing
curvature. This shows that there is a priori no relation with the Regge 
curvature \cite{Regg61} which is given at point 4 by 
$2 \pi - \th_{142}- \th_{143}- \th_{243}$. 
We will return to this example in the next section (see example 5 there).
                                                    \hfill $\blacksquare$

\section{Transformations to `local' tensor products and covariant tensor 
components}
\setcounter{equation}{0}
As in the preceding section, we consider a finite set $\M$ and a differential 
calculus $\O$ (over the algebra of functions) on $\M$.
In ordinary (continuum) differential geometry, the tensor product $\oa$ and the 
graded product in the space of differential forms are operations which take place
over the same point. This is not so in the discrete framework under consideration.
 For example, in $e^{ij}\oa e^{jk}$ the first factor is an element of
$\O^1_i$ while the second factor belongs to $\O^1_j$. In contrast, in 
$e^{ij}\otimes_L e^{ik}$ both factors belong to the same cotangent space.
As a consequence, the left components of an element of $\O^1 \otimes_L \O^1$
transform covariantly under a change of module basis in $\O^1$ (in contrast to the left,
middle or right components of an element of $\O^1 \oa \O^1$). Covariant tensor components
are of particular interest because of the possibility to construct new tensors from
them via contraction. For example, we would like to build a kind of Ricci tensor 
from the curvature components $R(i)^j{}_{klm}$ in (\ref{curv_comp}). The latter
are not covariant, however. The indices $j$ and $l$ (or $m$) live in different
(co)tangent spaces. In this section, we shall consider ways to modify or, more
precisely, to `localize' expressions in order to provide a remedy for this problem. 
What we need is tensor products which act over the same point and furthermore suitable
transformations from tensor products over $\A$ to these `local' tensor products.
Given a connection, we have the parallel transports which enable us to move from 
one (co)tangent space to another and these should be expected as natural ingrediences 
of the transformations we are looking for.
\vskip.2cm

A map $\O^1 \otimes_L \O^1 \rightarrow \O^1 \oa \O^1$ is given by
\be
   \kappa(\alpha \otimes_L \beta) 
   := \sum_{i,j} (e^i \, \alpha \, e^j) \oa [(e^i \, \beta) \U^{ij}]  \;.
\ee
In particular,
\be
   \kappa(e^{ij}\otimes_L e^{ik}) = \sum_l U(i)^k{}_{j l}\, e^{ij}\oa e^{jl} \; .
\ee
$\kappa$ is a left $\A$-homomorphism and has the property\footnote{This
shows that left $\A$-homomorphisms $\O^1 \otimes_L \O^1 \rightarrow
\O^1 \oa \O^1$ are in one-to-one correspondence with left $\A$-module 
linear connections.}
\be
    \kappa(\rho \otimes_L \beta) = \U(\beta)  \; .
\ee

A map 
\be
   \lambda_1 \, : \, \O^1 \oa \O^1 \rightarrow \O^1 \otimes_L\O^1
\ee
in the opposite direction is not so easily at hand in
an explicit form, except in some special cases like those listed below.
\begin{itemize}
\item 
If for all $i \longrightarrow j$ the transport $\U^{ij}$ is
invertible, we can define
\be
   \lambda_1(\alpha \oa \beta) := \sum_{i,j} (e^i \, \alpha \, e^j) \otimes_L
   [(e^j \, \beta) (\U^{ij})^{-1}]  \; .
\ee
Then $\lambda_1 = \kappa^{-1}$. This choice is considered in case of
the oriented lattice structures treated in sections 5 and 6.
\item
If the digraph associated with $\O^1$ is symmetric (i.e., a digraph where 
$i \longrightarrow j \Longleftrightarrow j \longrightarrow i$)  
then we may define\footnote{If $e^i \alpha \, e^j \neq 0$,
then $i \longrightarrow j$ with which the parallel transport $\U^{ij}$ is associated.
But instead, $\U^{ji}$ enters the above formula for $\lambda_1$.  Therefore
the symmetry condition is needed.}
\be       \label{lambda1_symm}
   \lambda_1(\alpha \oa \beta) := \sum_{i,j} (e^i \, \alpha \, e^j) \otimes_L
   [(e^j \, \beta) \U^{ji}]  \; .
\ee
\end{itemize}
In the following we assume that a map $\lambda_1$ is given, having the above
examples in mind. Moreover, we will also need a similar map
\be
   \lambda_2 \, : \, \O^2 \oa \O^1 \rightarrow \O^2 \otimes_L \O^1
\ee
(and furthermore a way to `localize' 2-forms, see below). 
In our examples considered in sections 5 and 6, $\lambda_1$ induces such a 
map $\lambda_2$ in a natural way.

\vskip.2cm
\noindent
{\em Example 1.}
Let  $i \longrightarrow j \longrightarrow k \longrightarrow l$ and 
$k \longrightarrow i$. For $e^{ijk} \neq 0$ we may define
\be
    \lambda_2(e^{ijk}\oa e^{kl}) := e^{ijk}\otimes_L [(e^{kl}) \U^{ki}]   \,  .
\ee
If also $k \longrightarrow j \longrightarrow i $, another choice is
\be
 \lambda'_2(e^{ijk}\oa e^{kl}) := e^{ijk}\otimes_L [(e^{kl}) \U^{kj} \U^{ji}]  \;.
\ee
The two choices for $\lambda_2$ can be different as long as the holonomy of the connection 
is not trivial. Hence, in general there are many different choices for $\lambda_2$.
                                              \hfill    $\blacksquare$

\vskip.2cm
\noindent
{\em Example 2.}
Let us now consider a differential calculus where the space of 1-forms is associated
with a symmetric digraph and let us moreover assume that the differential calculus
is basic (cf section 3.2). 
In this case, $e^{i_0 \cdots i_r} \neq 0$ implies that $i_k \longrightarrow i_l$ for all
$0 \leq k,l \leq r$ (cf \cite{Zapa97}). A natural choice for $\lambda_1$, $\lambda_2$
and generalizations thereof is then given by
\be     \label{lambda_symm_Z}
 \lambda(e^{i_0 \cdots i_r}\oa e^{i_r j}) :=
 e^{i_0\cdots i_r} \oa  [(e^{i_r j}) \U^{i_r i_0}]   \; .
\ee
                                              \hfill $\blacksquare$
\vskip.2cm

\noindent
In the following we simply write $\lambda$ instead of $\lambda_1$ or $\lambda_2$.
\vskip.2cm

Combining $\kappa$ and $\pi$,  
\be
   \alpha \cap \beta := \pi \circ \kappa (\alpha \otimes_L \beta)
\ee
determines a product $\O^1 \otimes_L \O^1 \rightarrow \O^2$ which is left
$\A$-linear and therefore satisfies 
\be
     e^i \, (\alpha \cap \beta) = (e^i \alpha) \cap (e^i \beta)  
\ee
so that $\cap$ preserves `locality'.
If $(\kappa \circ \lambda)(\ker \pi) \subset \ker \pi$,
the map 
\be     \label{mu_def}
   \mu := \pi \circ \kappa \circ \lambda \circ \pi^{-1} \, :  \, \O^2 \rightarrow \O^2
\ee
is well-defined and can be used to transform usual products of 1-forms (i.e., 
elements of $\O^2$) to $\cap$-products.

\vskip.2cm
\noindent
{\em Example 3.}
Let us again consider the case of a differential calculus associated with a
{\em symmetric} digraph.
Using (\ref{lambda1_symm}), we get
\be
  \kappa \circ \lambda_1(\alpha \oa \beta) &=& \sum_{i,j} 
  (e^i \alpha \, e^j) \oa [(e^j \beta) \H^{ji}]         \\
  \lambda_1 \circ \kappa (\alpha \otimes_L \beta) &=& \sum_{i,j} 
  (e^i \alpha \, e^j) \otimes_L [(e^i \beta) \H^{ij}]  
\ee
with the holonomies $\O^1_i \rightarrow \O^1_i$ given by $\H^{ij}$.
Then
\be
  \mu (\alpha \beta) = \sum_{i,j} (e^i \alpha \, e^j) 
            \cap [(e^j \beta) \U^{ji}]  
   = \sum_{i,j}(e^i \alpha) \, [(e^j \beta) \H^{ji}] \; .
\ee
The 2-form relations are of the form
\be 
   \sum_k e^{ikj}=0   \qquad  \mbox{if } \;  i  \, \not \! \! \longrightarrow j  
\ee
(where $k$ runs over a subset of $\M$) and must be mapped to $0$ by $\mu$. 
In terms of the $\cap$-product the 2-form relation then read
\be
  \sum_{k,l} U(k)^j{}_{il}\,e^{ik}\cap e^{il}= 0   \qquad
   \mbox{if } \;   i \, \not \! \! \longrightarrow j    \; .
\ee
Using $(e^{kj}) \H^{ki} =: \sum_l (\H^{ki})^j{}_l \, e^{kl}$, the condition
$(\kappa \circ \lambda)(\ker \pi) \subset \ker \pi$ amounts to
\be
    \sum_k (\H^{ki})^j{}_l \, e^{ikl} = 0   \qquad  \forall \ell  
\ee
and thus induces restrictions on the connection, in general.
                                              \hfill $\blacksquare$

\vskip.2cm
\noindent
{\em Lemma.} For a basic differential calculus  $(\O, \d)$ and a torsion-free
linear connection, we have
\be
  e^{ij} \cap e^{ij} &=&  -\sum_k e^{ijk}    \nonumber \\
  e^{ij}\cap e^{ik} &=& e^{ijk} \qquad \mbox{if } j \neq k
                  \label{cap_lemma}
\ee
and the map $\mu$ defined in (\ref{mu_def}) with $\lambda$ from 
(\ref{lambda1_symm}) satisfies
\be
 \mu(e^{iji}) &=&  -\sum_k e^{ij} \cap e^{ik}    \nonumber \\
 \mu(e^{ijk}) &=&  e^{ij} \cap e^{ik}  \qquad  \mbox{if } i \neq k   \; .
               \label{mu_lemma}
\ee
\vskip.1cm
\noindent
{\em Proof:} (\ref{cap_lemma}) follows from 
$$  e^{ij} \cap e^{ik} =  \sum_m U(i)^k{}_{jm} \, e^{ijm}  $$
together with (\ref{Uu}).  (\ref{mu_lemma}) results from
$$  \mu(e^{ijk}) =  e^{ij} \cap [ e^{jk} \U^{ji} ] 
                          = e^{ij} \cap \sum_m U(j)^k{}_{im} \, e^{im} 
                          = e^{ij} \cap \sum_m (\delta^k_m - \delta^k_i) \, e^{im}
$$ 
using again (\ref{Uu}).     
                                                     \hfill $\blacksquare$
\vskip.2cm

Now we have everything at hand to `localize' torsion and curvature and 
to define corresponding covariant components as follows,
\be
   \mu \circ \Theta(e^{ij}) &=:& \sum_{k,l} Q(i)^j{}_{kl} \, e^{ik} \cap e^{il}
  \\
 (\mu \otimes_L {\rm id}) \circ \lambda \circ \R(e^{ij}) &=:& \sum_{k,l,m} 
 \hat{R}(i)^j{}_{klm} \, (e^{il} \cap e^{im}) \otimes_L e^{ik} \; .
\ee
As in ordinary differential geometry, a {\em Ricci tensor} can now be
defined,
\be                    \label{Ricci_generalized}
 {\rm Ric}(i)_{jk} :=\sum_l \hat{R}(i)^l{}_{jl k} \, , \qquad
 \overline{\rm Ric}(i)_{jk} := \sum_l \hat{R}(i)^l{}_{jkl}  \; .
\ee
There is also the contraction $\sum_l \hat{R}(i)^l{}_{ljk}$ which in classical
Riemannian geometry vanishes identically. In the present framework its 
significance has still to be explored.
\vskip.2cm

In order to construct a curvature scalar, we need an inverse of $g(i)$. 
This need not exist at all vertices of the digraph. There are examples where 
$g(i)$ is even degenerate at all vertices.    

\vskip.2cm
\noindent
{\em Example 4.}
We continue our example 2. With the assumptions made there, there are no 
conditions on the connection (cf  example 3). For the curvature we obtain
\be
 (\mu \otimes_L {\rm id}) \circ \lambda \circ \R(e^{im}) =
 \sum_{i,j,k} e^{ij} \cap [e^{jk} \, \U^{ji}] \otimes_L \bigl\{ (e^{im})
 [\U^{ij} \U^{jk} \U^{ki} - \H^{ik}] \bigr\}
\ee
which for $e^{ij} \cap e^{ik} \neq 0$ yields
\be
   \hat{R}(i)^m{}_{n\,jk} = \sum_l U(j)^l{}_{ik} \, 
   [\U^{ij} \U^{jl} \U^{li} - \H^{il}]^m{}_n  \; .
\ee
                                         \hfill $\blacksquare$

\vskip.2cm
\noindent
{\em Example 5.}
We continue our example 4 of section 3.5.1 and choose $\lambda$ as in
(\ref{lambda_symm_Z}). 
The relations between the usual graded and the $\cap$-product are obtained
from the above Lemma. In particular,
\be
  e^{41} \cap e^{41} = -e^{412}-e^{413}-e^{414} \; , \qquad
  e^{41}\cap e^{42} = e^{412}
\ee
and
\be
  e^{12}\cap e^{13} = 0 \; , \qquad e^{12} \cap e^{12} = -e^{121}-e^{124} \; ,
  \qquad e^{12} \cap e^{14} = e^{124}  \; .
\ee
Since $\H^{ij} = I$, the map $\mu$ is well-defined. Then
\be
  \mu(e^{414}) = -e^{41} \cap e^{41}- e^{41} \cap e^{42}- e^{41} \cap e^{43}
  \; , \qquad \mu(e^{412}) = e^{41} \cap e^{42}
\ee
and
\be
  \mu(e^{123}) = e^{12}\cap e^{13} = 0 \, , \; \;
  \mu(e^{121}) = -e^{12} \cap e^{12}- e^{12} \cap e^{14} \, , \; \;
  \mu(e^{124}) = e^{12} \cap e^{14} . \;
\ee
The curvature $\hat{R}(i)_{jk} := (\hat{R}(i)^m{}_{njk})$ at point 4 is given by 
\be
      \hat{R}(4)_{11} = \hat{R}(4)_{22} =  \hat{R}(4)_{33} = 0
\ee
and
\be
 \hat{R}(4)_{12} = \hat{R}(4)_{21}&=& \left(\begin{array}{ccc}
      0 & 0 & 2 \, (ac-b)/(c^2-1) \\
      0 & 0 & 2 \, (bc-a)/(c^2-1) \\
      0 & 0 & -2 \end{array}\right)     \nonumber \\
 \hat{R}(4)_{13} = \hat{R}(4)_{31}&=&\left(\begin{array}{ccc}
      0 & 2 \, (ab-c)/(b^2-1) & 0 \\
      0 & -2              & 0 \\
      0 & 2 \, (bc-a)/(b^2-1) & 0 \end{array}\right)   \nonumber \\
 \hat{R}(4)_{23} = \hat{R}(4)_{32} &=& \left(\begin{array}{ccc}
      -2              & 0 & 0 \\
      2 \, (ab-c)/(a^2-1) & 0 & 0 \\
      2 \, (ac-b)/(a^2-1) & 0 & 0 \end{array}\right)  \; .
\ee 
 Furthermore, we have $\hat{R}(1)_{22} = \hat{R}(1)_{33} 
 = \hat{R}(1)_{44} = 0$, 
\be
 \hat{R}(1)_{24} = \hat{R}(1)_{42} = \left(\begin{array}{ccc}
      0 & 2 \, (bc-a)/(c^2-1) & 0 \\
      0 & -2              & 0 \\
      0 & 2 \, (1-a-b+c)/(c+1) & 0 \end{array}\right) 
\ee
etc. and corresponding expressions for the curvature at the vertices $2$
and $3$. For the Ricci tensors, we find ${\rm Ric}(i) = \overline{\rm Ric}(i)$,
\be
 {\rm Ric}(4) \!\! &=& \!\! 2  \left(\begin{array}{ccc}
       0 & (ac-b)/(a^2-1) & (ab-c)/(a^2-1) \\
       (bc-a)/(b^2-1) & 0 & (ab-c)/(b^2-1) \\
       (bc-a)/(c^2-1) & (ac-b)/(c^2-1) & 0 
  \end{array}\right)            \\
  {\rm Ric}(1) \!\! &=& \!\! 2 \left(\begin{array}{ccc}
            0 & (1-a+b-c)/(b+1) & (bc-a)/(b^2-1) \\
            (1-a-b+c)/(c+1) & 0 & (bc-a)/(c^2-1) \\
            0 & 0 & 0
   \end{array}\right)  \qquad
\ee
and corresponding expressions for ${\rm Ric}(j)$, $j=2,3$.
The resulting expression for the curvature scalar turns out to be rather complicated. 
In the special case $a=b=c$, we obtain
\be
   R(4) = \sum_{i,j} g(4)^{ij} {\rm Ric}(4)_{ij} =
   {1\over \ell^2} \, {12 \, a^2 \over (a-1)(a+1)(2a+1)}
\ee
and
\be
   R(1) = R(2) = R(3) =  - {1\over \ell^2}\, {8 \, a \over (a+1)(2a+1)} \; .
\ee
                                         \hfill $\blacksquare$
\vskip.2cm

The structures introduced in this section will also be exploited in the examples 
presented in the following two sections.

\section{Geometry of the oriented lattice}
\setcounter{equation}{0}
In this section we choose ${\cal M} = \mathbb{Z}^n = \{ a=(a^\mu) \, | \, 
a^\mu \in \mathbb{Z}\, , \, \mu=1,\ldots,n \}$ and consider the differential calculus
with
\be
 e^{ab} \neq 0  \qquad \Longleftrightarrow \qquad  b = a + \hat{\mu}
 \; \mbox{ for some } \mu                           \label{lrl} 
\ee
where $\hat{\mu} := (\delta^\nu_\mu) \in {\cal M}$. 
The corresponding graph is an oriented lattice in $n$ dimensions, 
a finite part of it is drawn in Figure 1.
Note that here we are dealing with an {\em infinite} set $\M$
for which in the formalism presented in the previous section in general technical 
problems associated with infinite sums arise. 
In the example under consideration we now sketch a transition to 
a formulation which then only makes reference to finitely generated
$\A$-modules so that only finite sums appear and it is safe
working on a purely algebraic level (see also \cite{DMH94_ddm}).

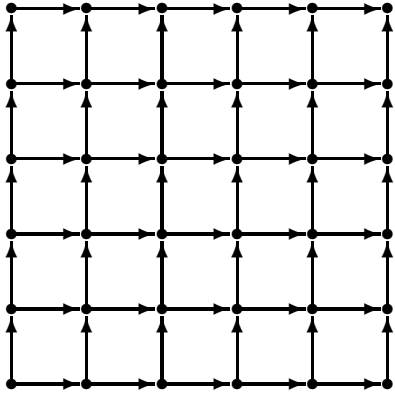
\begin{figure}
\centering
\unitlength1.cm
\begin{picture}(5.,5.2)(0.,0)
\thicklines
%\linethickness{0.3mm}
%
\multiput(0,0)(1,0){6}{\circle*{0.15}}
\multiput(0,1)(1,0){6}{\circle*{0.15}}
\multiput(0,2)(1,0){6}{\circle*{0.15}}
\multiput(0,3)(1,0){6}{\circle*{0.15}}
\multiput(0,4)(1,0){6}{\circle*{0.15}}
\multiput(0,5)(1,0){6}{\circle*{0.15}}
\multiput(0,0)(1,0){5}{\vector(1,0){0.9}}
\multiput(0,1)(1,0){5}{\vector(1,0){0.9}}
\multiput(0,2)(1,0){5}{\vector(1,0){0.9}}
\multiput(0,3)(1,0){5}{\vector(1,0){0.9}}
\multiput(0,4)(1,0){5}{\vector(1,0){0.9}}
\multiput(0,5)(1,0){5}{\vector(1,0){0.9}}
\multiput(0,0)(0,1){5}{\vector(0,1){0.9}}
\multiput(1,0)(0,1){5}{\vector(0,1){0.9}}
\multiput(2,0)(0,1){5}{\vector(0,1){0.9}}
\multiput(3,0)(0,1){5}{\vector(0,1){0.9}}
\multiput(4,0)(0,1){5}{\vector(0,1){0.9}}
\multiput(5,0)(0,1){5}{\vector(0,1){0.9}}

\end{picture}

\caption{\small A finite part of the oriented lattice graph.}
\end{figure}

\vskip.2cm
Each $f \in {\cal A}$ can be written as a function of 
\be
              x^\mu :=  \ell_\mu \, \sum_a a^\mu \, e^a
\ee
and its differential is then given by
\be
    \d f = \sum_\mu  \partial_{+\mu} f  \; \d x^\mu 
\ee
where
\be
   (\partial_{+ \mu}f)(x) 
 :=  {1 \over \ell_\mu} \, \lbrack f(x + \vec{\mu}) - f(x) \rbrack  
\ee
with $\vec{\mu}=\ell_\mu\,\hat{\mu}$.
The 1-forms $\d x^\mu$ constitute a basis of $\O^1$ as a 
left (or right) $\cal A$-module and satisfy the following commutation relations
with a function of $x^\mu$,
\be             \label{f-dx-latt}
    \d x^\mu \, f(x) = f(x + \vec{\mu}) \, \d x^\mu  \; .    
\ee
In particular, this implies 
\be
   \d x^\nu \bullet \d x^\mu =  [ \d x^\mu , x^\nu] 
  = \ell_\mu \, \delta^{\mu\nu} \, \d x^\mu
\ee
(cf also \cite{BDMH95}) and, acting with $\d$ on the latter equation, leads to 
\be              \label{dxdx_latt}
       \d x^\mu \, \d x^\nu + \d x^\nu \, \d x^\mu = 0   \; .
\ee
The 1-form $\rho$ introduced in (\ref{rho}) becomes
\be
      \rho = \sum_\mu {1 \over \ell_\mu} \,  \d x^\mu  \; .
\ee
It satisfies $\d \rho = 0$ and $\rho^2 = 0$. Moreover, for $w \in \O^r$ 
we have
\be
    \d w = \rho \, w - (-1)^r \, w \, \rho  \; .
\ee
\vskip.2cm

 For a linear (left $\A$-module) connection on $\O^1$ we write
\be
  \nabla \d x^\mu = - \sum_\nu \Gamma^\mu{}_\nu \oa \d x^\nu \, , \qquad
  \U(\d x^\mu) = \sum_\nu  U^\mu{}_\nu \oa \d x^\nu  \; .
\ee
Using (\ref{U_rho_nabla}) this leads to
\be                  \label{U_rho_Gamma}
   U^\mu{}_\nu  =   \rho \, \delta^\mu_\nu + \Gamma^\mu{}_\nu  
   =: \sum_\sigma {1 \over \ell_\sigma} \, U^\mu{}_{\sigma \nu}\, \d x^\sigma  \; .
\ee
We shall require that $\lim_{\ell \to 0} U^\mu_{\sigma \nu} =  \delta^\mu_\nu$.
This assumption will be used below where we work out continuum limits of
curvature expressions.
\vskip.2cm
 
The map $\kappa$ introduced in section 4 is given by
\be
   \kappa (\d x^\mu \otimes_L \d x^\nu) = \sum_\sigma U^\nu{}_{\mu \sigma}\, 
   \d x^\mu \oa \d x^\sigma  \; .
\ee
 For the left  $\A$-linear $\cap$-product in $\O^2$ we now obtain
\be
   \d x^\mu \cap \d x^\nu = \sum_\sigma U^\nu{}_{\mu \sigma}
   \, \d x^\mu \, \d x^\sigma \; .
\ee
Under a change of coordinates, $\d x^\mu \cap \d x^\nu$ transforms covariantly 
while $\d x^\mu \, \d x^\nu$ does not. Not all of the 2-forms $\d x^\mu \cap \d x^\nu$ 
are independent, in particular as a consequence of (\ref{dxdx_latt}). In the following we 
derive the relations which they satisfy under the assumption  that $\kappa$ has an
inverse which means that $U^\mu{}_\nu$ has an inverse 
$V^\mu{}_\nu = \sum_\sigma (1/ \ell_\sigma) \, V^\mu{}_{\sigma \nu}\, \d x^\sigma$ 
in the sense that
\be
  \sum_\sigma U^\mu{}_\sigma \bullet V^\sigma{}_\nu 
 =  \rho \, \delta^\mu_\nu 
 = \sum_\sigma V^\mu{}_\sigma \bullet U^\sigma{}_\nu  \; .
\ee
In terms of components this becomes    
\be
  \sum_\sigma U^\mu{}_{\alpha \sigma}\, V^\sigma{}_{\alpha \nu} 
 = \delta^\mu_\nu  
 =  \sum_\sigma V^\mu{}_{\alpha \sigma}\, U^\sigma{}_{\alpha \nu}  
\ee
for all  $\alpha$. Now we have
\be 
  \d x^\mu \, \d x^\nu = \sum_\sigma V^\nu{}_{\mu \sigma}\, \d x^\mu 
                                  \cap \d x^\sigma  \; .
\ee 
We introduce
\be
     W^{\mu \nu}_{\rho \sigma} := U^\nu{}_{\mu \rho}\, V^\mu{}_{\rho \sigma}
\ee
which satisfies $\lim_{\ell \to 0} W^{\mu\nu}_{\rho\sigma} = 
\delta^\mu_\sigma \, \delta^\nu_\rho$ and 
\be
    \sum_{\kappa,\lambda} W^{\mu \nu}_{\kappa \lambda} \,  
   W^{\kappa \lambda}_{\rho \sigma} = \delta^\mu_\rho \, \delta^\nu_\sigma  \; .
\ee
As a consequence,
\be
   (P^\pm)^{\mu \nu}_{\rho \sigma} := {1 \over 2} \, ( \delta^\mu_\rho \, \delta^\nu_\sigma
   \pm W^{\mu \nu}_{\rho \sigma} )
\ee
are projectors. In terms of the $\cap$-product, the 2-form relations (\ref{dxdx_latt}) can
now be expressed as follows,
\be          \label{cap_comm_rels}
 \sum_{\kappa,\sigma} (P^+)^{\mu \nu}_{\kappa \sigma} \,
   \d x^\kappa \cap \d x^\sigma = 0    \; .
\ee
This much more complicated form of the 2-form relations, as compared with
(\ref{dxdx_latt}), is the price we have to pay for the covariance. 
 For a 2-form $A =  \sum_{\mu,\nu} A_{\mu \nu} \, \d x^\mu \, \d x^\nu
 =  \sum_{\mu,\nu} \hat{A}_{\mu \nu} \, \d x^\mu \cap \d x^\nu$ we
obtain the implications  
\be  
 A = 0   \quad   \Longleftrightarrow \quad  
  \sum_{\kappa, \sigma}
  (P^-)^{\kappa \sigma}_{\mu \nu} \,  \hat{A}_{\kappa \sigma} = 0
\ee
and
\be
 A_{\mu \nu}+A_{\nu \mu} = 0 \quad \Longleftrightarrow \quad
 \sum_{\kappa, \sigma} (P^+)^{\kappa \sigma}_{\mu \nu} \, 
     \hat{A}_{\kappa \sigma} = 0
\ee
(since $A_{\mu \nu} = \sum_\rho \hat{A}_{\mu \rho}\, U^\rho{}_{\mu\nu}$). 
\vskip.2cm

With the help of (\ref{U_rho_Gamma}), our general expression (\ref{torsion}) for 
the torsion of a linear connection leads to
\be
   \Theta^\mu &:=& \Theta(\d x^\mu) = \sum_{\nu,\rho} {1 \over \ell_\nu} 
       (U^\mu{}_{\nu\rho}- \delta^\mu_\rho) \, \d x^\nu \, \d x^\rho
       \nonumber \\
 &=& \sum_{\nu,\rho,\sigma} {1 \over \ell_\nu} (U^\mu{}_{\nu\rho}- 
    \delta^\mu_\rho) \,V^\rho{}_{\nu \sigma}\, \d x^\nu \cap \d x^\sigma  \; .
\ee
Writing
\be
   \Theta^\mu = {1\over 2} \, \sum_{\nu,\rho}Q^\mu{}_{\nu \rho}\, 
                         \d x^\nu \cap \d x^\rho
\ee
where the coefficients $Q^\mu{}_{\nu \rho}$ are subject to
\be 
  Q^\mu{}_{\nu\rho} = - \sum_{\kappa,\lambda} W^{\kappa \lambda}_{\nu\rho}
  \, Q^\mu{}_{\kappa \lambda} \, ,
\ee
we are led to
\be
    Q^\mu{}_{\nu\rho} = \sum_{\kappa,\lambda,\sigma}{1 \over \ell_\kappa}\,
       (\delta^\kappa_\nu \delta^\lambda_\rho - W^{\kappa\lambda}_{\nu\rho})
   \, (U^\mu{}_{\kappa \sigma}- \delta^\mu_\sigma) \, 
   V^\sigma{}_{\kappa\lambda} \; .
\ee
\vskip.2cm
\noindent
{\em Example.} If the torsion vanishes, we obtain
\be
 {1 \over \ell_\nu}(U^\mu{}_{\nu \rho}- \delta^\mu_\rho) =
 {1\over \ell_\rho}(U^\mu{}_{\rho \nu}- \delta^\mu_\nu)   \; .
\ee
This is equivalent to the condition
\be
     \Gamma^\mu{}_{\nu \rho} = \Gamma^\mu{}_{\rho \nu}
\ee
which is familar from continuum differential geometry. 
                                                 \hfill $\blacksquare$
\vskip.2cm

A metric tensor (in the sense of section 3) is given by 
\be
     g = \sum_{\mu,\nu} g_{\mu \nu} \, \d x^\mu \otimes_L \d x^\nu
\ee
where $g_{\mu \nu}$ is now assumed to be a non-degenerate symmetric matrix. 
The metric compatibility condition $\nabla g = 0$ with a linear connection 
$\nabla$ leads to
\be
    g(x + \vec{\lambda})_{\mu \nu} 
 = \sum_{\rho,\sigma}U(x)^\rho{}_{\lambda \mu} \,
    g(x)_{\rho \sigma} \, U(x)^\sigma{}_{\lambda \nu}
\ee
for all $\lambda$.  In matrix notation, this takes the form
\be
   g(x+\vec{\lambda}) = U(x)_\lambda^t \, g(x) \, U(x)_\lambda  \; .
\ee
The continuum limit of this equation is obtained from the 
expansion
\be
  & & \tilde{g}_{\mu\nu} + \ell_\lambda (\pa_\lambda \tilde{g} _{\mu \nu} 
   + b_{\mu\nu}) + {\cal O}(\ell_\mu^2)             \nonumber \\
 &=& \sum_{\rho,\sigma} (\delta^\rho_\mu + \ell_\lambda \, 
    \Gamma^\rho{}_{\lambda\mu}) \, 
    g_{\rho \sigma} \, (\delta^\sigma_\nu + \ell_\lambda \, 
   \Gamma^\sigma{}_{\lambda\nu})                     \nonumber \\
 &=& \tilde{g}_{\mu \nu} + \ell_\lambda \, \big( \sum_{\rho} 
   (\tilde{\Gamma}^\rho{}_{\lambda\mu} \, 
   \tilde{g}_{\rho\nu}+ \tilde{g}_{\mu \rho} \, 
   \tilde{\Gamma}^\rho{}_{\lambda \nu}) +b_{\mu\nu} \big) 
   + {\cal O}(\ell_\mu^2)   \qquad
\ee
where 
\be					
   \tilde{\Gamma}^\mu{}_{\sigma \nu} := \lim_{\ell_\lambda \to 0} 
   \Gamma^\mu{}_{\sigma \nu}                                               \; , \qquad
   \tilde{g}_{\mu\nu}:=\lim_{\ell_\lambda \to 0} g_{\mu\nu}     \; , \qquad
   b_{\mu\nu} := \lim_{\ell_\lambda \to 0} {\pa g_{\mu\nu}\over \pa \ell_\mu}
\ee
which we assume to exist.  

\vskip.2cm
\noindent
{\em Remark.}
The vector fields $\pa_{+\mu} \in \X$ are dual to the 1-forms $\d x^\mu$, i.e.,
\be
    \langle \d x^\mu , \pa_{+\nu} \rangle = \delta^\mu_\nu   \; .
\ee
The action of $X = \sum_\mu \pa_{+\mu} \cdot X^\mu$ on functions is given by
\be
    X(f) = \langle \d f, X \rangle = \sum_\mu  X^\mu \, (\pa_{+\mu} f)  \; .
\ee
 For the connection we have  $\U(X) = X \oa \rho + \nabla X$ and thus
\be
   \U(\pa_{+\mu}) = \sum_\nu \, \pa_{+\nu}\oa U^\nu{}_\mu \; .
\ee
A dual metric tensor (cf section 3) can be expressed as
\be
     h = \sum_{\mu,\nu} \pa_{+\mu}\otimes_R \pa_{+\nu}\cdot h^{\mu\nu}
\ee
with components $h^{\mu \nu} \in \A$. The metric compatibility condition for
a linear connection takes the form $\U(h) = h \oa \rho$. The latter leads to
\be
  h(x+\vec{\lambda})^{\mu\nu} = \sum_{\rho,\sigma}
  V(x)^\mu{}_{\lambda\rho}\, V(x)^\nu{}_{\lambda\sigma}\, 
  h(x)^{\rho \sigma} \; .
\ee
With $h^{\mu \nu} = g^{\mu \nu}$, where $g^{\mu \nu}$ are the components
of the matrix inverse to $(g_{\mu \nu})$, we obtain the metric tensor inverse
to $g$.
                                    \hfill $\blacksquare$
\vskip.2cm

Let us now turn to the calculation of the curvature of a linear connection. We have 
\be 
  & & \R(\d x^\mu) = \sum_\nu ( \d \Gamma^\mu{}_\nu + \sum_\rho \Gamma^\mu{}_\rho
  \, \Gamma^\rho{}_\nu ) \oa \d x^\nu                                  \nonumber \\ 
  &=&  \sum_{\rho,\nu} U^\mu{}_\rho \, U^\rho{}_\nu
   \oa \d x^\nu = \sum_{\rho,\kappa,\lambda,\nu} {1 \over \ell_\kappa\ell_\lambda} 
   U(x)^\mu{}_{\kappa \rho} \, U(x+\vec{\kappa})^\rho{}_{\lambda\nu}
   \, \d x^\kappa \, \d x^\lambda \oa \d x^\nu   \nonumber \\
 &=& {1 \over 2} \, \sum_{\kappa,\lambda,\nu}{1 \over \ell_\kappa\ell_\lambda}
  [U(x)_\kappa U(x+\vec{\kappa})_\lambda - U(x)_\lambda U(x+
  \vec{\lambda})_\kappa]^\mu{}_\nu \, \d x^\kappa \, \d x^\lambda \oa \d 
  x^\nu \; .
\ee
With 
\be
   \R(\d x^\mu) =: {1 \over 2} \,  \sum_{\kappa,\lambda,\nu} R^\mu{}_{\nu \kappa
 \lambda} \, \d x^\kappa \, \d x^\lambda \oa \d x^\nu 
\ee
where $ R^\mu{}_{\nu \kappa \lambda} = - R^\mu{}_{\nu \lambda \kappa}$, we thus 
have
\be
  R^\mu{}_{\nu \kappa \lambda} = {1 \over \ell_\kappa \ell_\lambda}
  [U(x)_\kappa U(x+\vec{\kappa})_\lambda - U(x)_\lambda 
  U(x+ \vec{\lambda})_\kappa]^\mu{}_\nu  \; .
\ee
To obtain the tensorial components of the curvature, we need to transform $\oa$
into $\otimes_L$ and the $\d x^\kappa \, \d x^\lambda$ into
$\d x^\kappa \cap \d x^\lambda$. We achieve this with 
$\lambda = \kappa^{-1}$. First we note that
\be
    \lambda (\d x^\mu \oa \d x^\nu) = 
    \sum_\rho V(x)^\nu{}_{\mu \rho} \, \d x^\mu \otimes_L \d x^\rho
\ee
and therefore\footnote{The intermediate result in the second line is not
well-defined, but helps to understand how the final formula is obtained.}
\be
 & &  \lambda (\d x^\mu \, \d x^\nu \oa \d x^\rho)  \nonumber \\
 &=&
 {1\over2} \d x^\mu \, (\sum_\lambda V(x)^\rho{}_{\nu \lambda} \, 
  \d x^\nu \otimes_L 
  \d x^\lambda)-{1\over2} \d x^\nu \, (\sum_\lambda V(x)^\rho{}_{\mu \lambda} 
  \, \d x^\mu \otimes_L  \d x^\lambda)         \nonumber \\ 
  &=& {1\over2}\sum_{\lambda,\sigma}\left\{ V(x)^\lambda{}_{\mu\sigma}
          \d x^\mu \, (V(x)^\rho{}_{\nu \lambda} \, \d x^\nu) 
          - V(x)^\lambda{}_{\nu\sigma}
          \d x^\nu \, (V(x)^\rho{}_{\mu \lambda} \, \d x^\mu)
           \right\}
          \otimes_L \d x^\sigma                         \nonumber \\
  &=&{1\over2}\sum_{\sigma} [V(x+ \vec{\mu})_\nu \, V(x)_\mu +
           V(x+ \vec{\nu})_\mu \, V(x)_\nu]^\rho{}_\sigma \,
           (\d x^\mu \, \d x^\nu) \otimes_L \d x^\sigma  \; .
\ee
Applying this formula, we find
\be
 & & \lambda \circ \R(\d x^\mu) = {1 \over 4} \sum_{\kappa,\lambda,\nu}
    {1\over\ell_\kappa\ell_\lambda}\left[ \left( U(x)_\kappa \, 
    U(x+\vec{\kappa})_\lambda 
    - U(x)_\lambda \, U(x+ \vec{\lambda})_\kappa \right)\right. \nonumber  \\ 
  & & \times \,\left. \left( V(x+\vec{\kappa})_\lambda \,
     V(x)_\kappa + V(x+\vec{\lambda})_\kappa \,
     V(x)_\lambda \right) \right]^\mu{}_\nu\,    
     \d x^\kappa \, \d x^\lambda \otimes_L \d x^\nu \; .
\ee
With
\be
     \lambda \circ \R(\d x^\mu) =: \sum_\nu \hat{R}^\mu{}_\nu \otimes_L \d x^\nu
\ee
this leads to
\be
   \hat{R}^\mu{}_\nu  = {1 \over 4} \sum_{\kappa,\lambda}
   {1\over\ell_\kappa \ell_\lambda}
   \left[ H(x)_{\kappa \lambda}-H(x)_{\lambda \kappa}\right]^\mu{}_\nu 
   \, \d x^\kappa \, \d x^\lambda
\ee
where 
\be
 H(x)_{\kappa \lambda}:= U(x)_\kappa \, U(x+\vec{\kappa})_\lambda \, 
  V(x+ \vec{\lambda})_\kappa  V(x)_\lambda   \; .
\ee
Expressing the 2-forms $\hat{R}^\mu{}_\nu$ as follows,
\be
   \hat{R}^\mu{}_\nu = {1 \over 2} \, \sum_{\rho,\sigma} 
   \hat{R}^\mu{}_{\nu\rho\sigma} \, \d x^\rho \cap \d x^\sigma
\ee
with tensorial coefficients subject to
\be
   \hat{R}^\mu{}_{\nu \rho \sigma} = - \sum_{\kappa,\lambda} 
  W^{\kappa \lambda} _{\rho \sigma} \,  \hat{R}^\mu{}_{\nu \kappa \lambda} \, ,
\ee
we get
\be
   \hat{R}^\mu{}_{\nu \kappa \lambda} = {1 \over 2} \, 
   \sum_{\alpha}{1 \over \ell_\kappa \ell_\alpha}
   \left[ H(x)_{\kappa \alpha}-H(x)_{\alpha \kappa}\right]^\mu{}_\nu 
   \, V(x)^\alpha{}_{\kappa \lambda} \; .
\ee
The resulting Ricci tensors are
\be
   \mbox{Ric}_{\mu \nu} &=& {1 \over 2} \, 
   \sum_{\alpha, \beta}{1 \over \ell_\alpha \ell_\beta}
   \left[ H(x)_{\beta \alpha}- H(x)_{\alpha \beta}\right]^\beta{}_\mu 
   \, V(x)^\alpha{}_{\beta \nu} \\
  \overline{\mbox{Ric}}_{\mu \nu} &=& {1 \over 2} \, 
   \sum_{\alpha, \beta}{1 \over \ell_\alpha \ell_\nu}
   \left[ H(x)_{\nu \alpha}- H(x)_{\alpha \nu}\right]^\beta{}_\mu 
   \, V(x)^\alpha{}_{\nu \beta}
\ee
from which one obtains the curvature scalars 
$\hat{R} = g^{\mu \nu} \, \mbox{Ric}_{\mu \nu}$ and
$\overline{\hat{R}} = g^{\mu \nu} \, \overline{\mbox{Ric}}_{\mu \nu}$ 
with the help of the inverse $g^{\mu \nu}$ of $g_{\mu \nu}$. 
\vskip.2cm

In order to elaborate the continuum limit of the curvature tensor, we 
use the expansions
\be
  U(x)_\kappa &=& I + \ell_\kappa \,\tilde{\Gamma}_\kappa + {\ell_\kappa^2 \over 2} 
   \, \lbrack (\tilde{\Gamma}_\kappa)^2 + B_\kappa \rbrack + {\cal O}(\ell^3)  \\
  U(x+\vec{\lambda})_\kappa &=& I+ \ell_\kappa \, \tilde{\Gamma}_\kappa 
  + \ell_\kappa \ell_\lambda \, \pa_\lambda \tilde{\Gamma}_\kappa
  + {\ell_\kappa^2 \over 2} \, \lbrack (\tilde{\Gamma}_\kappa)^2 + B_\kappa \rbrack 
  + {\cal O}(\ell^3)   \\
   V(x)_\kappa &=& I - \ell_\kappa \,\tilde{\Gamma}_\kappa + 
   {\ell_\kappa^2 \over 2} \, \lbrack (\tilde{\Gamma}_\kappa)^2 - B_\kappa \rbrack
   + {\cal O}(\ell^3)  \\
  V(x+\vec{\lambda})_\kappa &=& I- \ell_\kappa \, \tilde{\Gamma}_\kappa 
  -\ell_\kappa\ell_\lambda\,\pa_\lambda\tilde{\Gamma}_\kappa 
   + {\ell_\kappa^2 \over 2} \, \lbrack (\tilde{\Gamma}_\kappa)^2 - B_\kappa \rbrack  
  + {\cal O}(\ell^3)  \;.
\ee
This leads to 
\be
  H(x)_{\kappa\lambda} = I + \ell_\kappa\ell_\lambda \, 
  [\pa_\kappa \tilde{\Gamma}_\lambda + \tilde{\Gamma}_\kappa 
  \tilde{\Gamma}_\lambda - \pa_\lambda \tilde{\Gamma}_\kappa 
  + \tilde{\Gamma}_\lambda \tilde{\Gamma}_\kappa] + {\cal O}(\ell^3)
\ee
so that
\be
   \hat{R}^\mu{}_{\nu \kappa \lambda} =  
  \pa_\kappa \tilde{\Gamma}^\mu{}_{\lambda \nu}
  - \pa_\lambda \tilde{\Gamma}^\mu{}_{\kappa \nu} + 
  \tilde{\Gamma}^\mu{}_{\kappa \rho} \tilde{\Gamma}^\rho{}_{\lambda \nu} 
  - \tilde{\Gamma}^\mu{}_{\lambda \rho} \tilde{\Gamma}^\rho{}_{\kappa \nu}
  + {\cal O}(\ell)  \; .
\ee
In this way we thus recover the continuum Riemann tensor in the limit $\ell \to 0$.
\vskip.2cm

We have set up a formalism which assigns geometrical notions like metric, curvature
and Ricci tensor to a hypercubic lattice. In particular, one obtains a discrete counterpart
of the Einstein (vacuum) equations in this way. Actually, there are several discrete Einstein
equations depending on our choice of Ricci tensor. The results of the following section 
suggest that the difference ${\rm Ric}-\overline{\rm Ric}$ is the appropriate object.

\vskip.2cm
\noindent
{\em Remark.}
The maps $\kappa$ and $\lambda$ extend to an arbitrary number of
factors of the corresponding tensor products. We define
\be
  \kappa(\alpha_1 \otimes_L \cdots \otimes_L\alpha_r) :=
 ({\rm id} \oa \kappa) \, [\alpha_1 \bullet \U(\alpha_2 \otimes_L \cdots
 \otimes_L \alpha_r)]
\ee
and correspondingly for $\lambda$. These maps allow us to introduce
covariant components of higher order forms by expressing them in terms of 
\be
   \alpha_1 \cap \cdots \cap \alpha_r := \pi \circ \kappa (\alpha_1 \otimes_L
   \cdots \otimes_L \alpha_r)  \; .
\ee
These $r$-forms satisfy very complicated relations which generalize (\ref{cap_comm_rels})
and involve the curvature, in general.
                                        \hfill $\blacksquare$

\section{Discrete surfaces of revolution}
\setcounter{equation}{0}
In terms of coordinates $\th,\ph$ we consider the differential calculus 
determined by
\be
  \d \th \, f(\th,\ph) = f(\th+\ell,\ph) \, \d \th \, , \qquad
  \d \ph \, f(\th,\ph) = f(\th,\ph+\ell) \, \d \ph  \; .
\ee
This is just a special case of (\ref{f-dx-latt}).  
Via the rules of differential calculus it leads to 
\be
  \d \th \, \d \th = 0 \, ,  \qquad 
  \d \th \, \d \ph + \d \ph \, \d \th = 0 \, ,  \qquad
  \d \ph \, \d \ph = 0   \; .
\ee
In contrast to the previous section, we interpret the coordinates as
spherical coordinates where $\th \in [0,\pi)$ and $\ph \in [0, 2 \pi)$.
With $\ell = \pi/n$, $n \in \mathbb{N}$, we obtain a discretization
of the surface by fixing one point on the surface and moving in steps
of coordinate length $\ell$ in $\th$- and $\ph$-directions. 
 For the metric we make an ansatz
\be 
   g(\th,\ph) = \left(\begin{array}{cc} 1 & 0\\0 & b^2
                     \end{array}\right)
\ee
where $b$ is a function of $\th$ only. This models a surface
of revolution (for example, a sphere as in Figure 2).
\vskip.2cm

Using $B := {\rm diag}(1,b)$, we have $g = B^t B$ and the metric compatibility 
condition for the parallel transport takes the form
\be
  (B U_\th \tilde{B}^{-1})^t \, (B U_\th \tilde{B}^{-1}) = I 
     \; ,  \qquad 
  (B U_\ph B^{-1})^t \, (B U_\ph B^{-1}) = I
\ee
where $\tilde{B} := {\rm diag}(1,\tilde{b})$ and $\tilde{b}(\th) :=
b(\th+\ell)$. As a consequence of these equations, $\tilde{B} \, U_\th \, B^{-1}$ 
and $B \, U_\ph \, B^{-1}$ are elements of the orthogonal group $O(2)$. 
In order to obtain the correct continuum limit, we restrict them to be
elements of $SO(2)$, the component of $O(2)$ which contains the
identity. Then we have expressions
\be
    U_\th = B^{-1} \, T(u) \, \tilde{B} \, ,\qquad
    U_\ph = B^{-1} \, T(v) \, B
\ee
where $u,v$ are arbitrary functions of $\th$ and $\ph$ and 
\be
  T(\chi) = \left(\begin{array}{cc} \cos \chi & -\sin \chi\\
                                                    \sin \chi & \cos \chi 
                       \end{array}\right)  \; .
\ee
The metric compatibility condition now leads to
\be
    U_\th = \left(\begin{array}{cc} 
                 \cos u & -\tilde{b} \sin u \\
                 (1/b) \sin u & (\tilde{b}/b) \cos u
                 \end{array}\right) \, ,  \qquad
    U_\ph = \left(\begin{array}{cc} 
                   \cos v & -b \, \sin v \\
                  (1/b) \sin v &  \cos v
                  \end{array}\right)
\ee
and the condition of vanishing torsion becomes
\be 
    \tilde{b} \sin u + \cos v = 1 \, ,  \qquad
    \tilde{b} \cos u - \sin v = b  \; .
\ee
These equations determine $u$ and $v$ completely in terms of $b$ and $\tilde{b}$.
We find 
\be
    \cos u = {1-p^2 \over 1+p^2} \, , \; 
    \sin u = {2 \, p \over 1+p^2} \, , \;
    \cos v = {1+p^2-2 \, \tilde{b}\, p \over 1+p^2} \, , \;
    \sin v = {\tilde{b}-b-(\tilde{b}+b) \, p^2 \over1+p^2}
\ee
with
\be
   p =  \left( 2 \, \tilde{b} \pm \sqrt{4 \, \tilde{b}^2-(\tilde{b}^2-b^2)^2} \right) 
          / (b+\tilde{b})^2  \; .
\ee
Only with the minus sign in the last expression we obtain a reasonable continuum
limit, and this choice will be made in the following.
The inverse parallel transport matrices are given by $V_\th = \tilde{B}^{-1}
T(-u) B$ and $V_\ph = B^{-1} T(-v) B$, so that
\be
     V_\th = \left(\begin{array}{cc} 
                  \cos u & b \sin u \\
                  - (1/\tilde{b}) \sin u & (b/\tilde{b}) \cos u
                  \end{array}\right) \, ,   \qquad
     V_\ph = \left(\begin{array}{cc} 
                    \cos v & b \sin v \\
                   -(1/b) \sin v &  \cos v
                   \end{array}\right)   \; .
\ee
With $\lambda = \kappa^{-1}$ (see section 4), we obtain for the curvature
\be
   \lambda \circ \R(\d x^\mu) = \sum_\nu r^\mu{}_\nu \, \d \th \, 
                                               \d \ph \otimes_L \d x^\nu 
\ee
where $x^1=\th, \, x^2=\ph$ and
\be
    r  &:=& {1 \over 2 \ell^2} [U_\th(\th,\ph) \, U_\ph(\th+\ell,\ph) \,
    V_\th(\th,\ph+\ell) \, V_\ph(\th,\ph)    \nonumber \\
   & &  - U_\ph(\th,\ph) \, U_\th(\th,\ph+\ell) \, V_\ph(\th+\ell,\ph) \, V_\th(\th,\ph)]  
                   \nonumber \\
   & =& {1 \over 2\ell^2} \, B^{-1} [T(u) \, T(\tilde{v}) \, T(-u) \, T(-v)-
             T(v) \, T(u) \, T(-\tilde{v}) \, T(-u)] \, B       \nonumber \\
   & =& {1 \over 2 \ell^2} \, B^{-1} \, [T(\tilde{v}-v)-T(v-\tilde{v})] \, B 
             \nonumber \\
   &=&     {1\over\ell^2} \, \left(\begin{array}{cc}
                                        0 & -b \, \sin(\tilde{v}-v) \\
                                        (1/b) \sin(\tilde{v}-v) & 0 
                                              \end{array}\right)    
\ee    
with $\tilde{v}(\th) := v(\th+\ell)$. Since $u$ and $v$ are functions of 
$b$ and $\tilde{b}$, they are functions of $\th$ only. Using 
\be
  \d \th \, \d \ph = V^\ph_{\th \th}\, \d \th \cap \d \th +
     V^\ph_{\th \ph}\, \d \th \cap \d \ph  \, ,  \quad
  \d \ph \, \d \th = V^\th_{\ph \th}\, \d \ph \cap \d \th +
  V^\th_{\ph \ph}\, \d \th \cap \d \ph
\ee
and $ r \, \d \th \, \d \ph = {1\over2}\, r \, (\d \th \, \d \ph - \d \ph \, \d \th)$,
we find the curvature components
\be
  \hat{R}_{\th \th} = - {\sin u\over \tilde{b}}\, r \, , \quad
  \hat{R}_{\th \ph} = {b \, \cos u \over \tilde{b}}\, r  \, , \quad
  \hat{R}_{\ph \th} = -(\cos v)  \, r  \, , \quad
  \hat{R}_{\ph \ph} = -b \, (\sin v) \, r  
\ee
where $\hat{R}_{\kappa \lambda} = (\hat{R}^\mu{}_{\nu \kappa \lambda})$.
We have the two Ricci tensors  
\be
{\rm Ric}&=& {1 \over \ell^2} \left(\begin{array}{cc} -(1/b) \cos v & -\sin v \\
                                                (b/\tilde{b}) \sin u & -(b^2/\tilde{b}) \cos u 
                                                 \end{array}\right) \, \sin(\tilde{v}-v)    \\
 \overline{{\rm Ric}} &=& {1\over\ell^2}
      \left(\begin{array}{cc} (1/\tilde{b}) \cos u & -\sin v \\
                                         (b/\tilde{b}) \sin u & b \, \cos v 
             \end{array}\right) \, \sin(\tilde{v}-v)  
\ee
and the combination 
\be     \label{Ric_comb}
  \widetilde{{\rm Ric}} := {1 \over 2} ({\rm Ric}-\overline{{\rm Ric}})
   = -{1 \over 2 \ell^2}({\cos u \over \tilde{b}}+{\cos v \over b})
       \, \sin(\tilde{v}-v) \; g   
\ee
from which we obtain the curvature scalars\footnote{The geometrically
interesting condition of a {\em constant} curvature scalar translates into a 
complicated difference equation for $b(\th)$, 
$$ \left[ {\cos u(\th) \over b(\th+\ell)} + {\cos v(\th) \over b(\th)} \right] 
     \, \sin[v(\th+\ell) - v(\th)] = \mbox{const.} $$
where $ b(\th+\ell) \, \sin u(\th) + \cos v(\th) = 1$ and
$ b(\th+\ell) \, \cos u(\th) - \sin v(\th) = b(\th)$.}
\be
    \hat{R} &:=& g^{\mu \nu}\, {\rm Ric}_{\mu \nu} 
                    = -{1 \over \ell^2} ({\cos u \over \tilde{b}}+{\cos v \over b})
                        \, \sin(\tilde{v}-v)   \\
    \bar{\hat{R}} &:=& g^{\mu \nu} \, \overline{\rm Ric}_{\mu \nu} 
                              = - \hat{R}            \\
    \widetilde{R} &:=&  g^{\mu \nu} \, \widetilde{\rm Ric}_{\mu \nu} =   \hat{R}  \; .
\ee
Now (\ref{Ric_comb}) becomes
\be                    \label{discr_Einstein_space}
    \widetilde{\rm Ric}_{\mu \nu} = {1 \over 2} \, \widetilde{R} \, g_{\mu \nu} \; .
\ee
These results clearly distinguish the particular linear combination (\ref{Ric_comb}) of
Ricci tensors.
\vskip.2cm

In the following, we present expansions in powers of $\ell$ and consider the
continuum limit $\ell \to 0$. We shall allow an explicit dependence of $b$ on $\ell$,
i.e., $b(\th,\ell) = b_0(\th) + b_1(\th) \, \ell + {\cal O}(\ell^2)$. Then
\be
  \Gamma_\th &=&  {1 \over \ell} \, (U_\th-I)     \nonumber \\
     &=&   \left(\begin{array}{cc}
    0 & 0 \\ 0 & b_0'/b_0 \end{array} \right) +
    \left(\begin{array}{cc}  0 & - (b_0'{}^2/2)   \\
    b_0'{}^2/2b_0^2  &  (2 b_1' + b_0'')/2b_0 - b_1 b_0'/b_0^2
    \end{array}\right) \, \ell + {\cal O}(\ell^2)       \nonumber \\
  \Gamma_\ph  &=&  {1 \over \ell} \, (U_\ph-I)   \nonumber \\
    &=&   \left(\begin{array}{cc}
   0 & -b_0 b_0' \\ b_0'/b_0 & 0  \end{array}\right) + \left( \begin{array}{cc} 
   -b_0'{}^2/2  &  -[b_1 b_0'+b_0 b_1' + b_0 b_0''/2 ] \\ 
   (2b_1'+b_0'')/2 b_0 -b_1 b_0'/ b_0^2 & -b_0'{}^2/2 
   \end{array}\right) \, \ell    \nonumber \\
  & &  + {\cal O}(\ell^2)
\ee
where $b'$ denotes the derivative of $b$ with respect to $\th$. 
 For the curvature, we find $\hat{R}_{\th \th} = {\cal O}(\ell^2)$ and
\be
   \hat{R}_{\th \ph} &=& \left(\begin{array}{cc}
    0 & -b_0 b_0'' \\ b_0''/b_0 & 0 \end{array}\right) \nonumber \\
   & & + \left(\begin{array}{cc}
   0 &  (-b_1+b_0') \, b''_0 - b_0 \, (b_1''+b_0''')   \\
   - [(b_1+b_0')b_0''/b_0^2-(b_1''+b_0''')/b_0] & 0
   \end{array}\right)\, \ell  \nonumber \\
   & & + {\cal O}(\ell^2)         \nonumber  \\
   \hat{R}_{\ph\th}&=& \left(\begin{array}{cc}
   0 & b_0 \, b_0' \\ - b_0''/b_0 & 0 \end{array}\right) \nonumber \\
   & & + \left(\begin{array}{cc}
   0 & b_1 b_0''+b_0(b_1''+b_0''')    \\
   -[b_1 b_0''-b_0(b_1''+b_0''')]/b_0^2  & 0
   \end{array}\right) \, \ell + {\cal O}(\ell^2)     \nonumber \\
   \hat{R}_{\ph\ph}&=& \left(\begin{array}{cc}
   0 & b_0^2 \, b_0' \, b_0''   \\
   - b_0' \, b_0''  & 0 \end{array}\right)\, \ell + {\cal O}(\ell^2)  \; .
\ee 
The Ricci tensors have the following expansions,
\be
  {\rm Ric}&=& \left(\begin{array}{cc}
  - b_0''/b_0 & 0 \\ 0 & - b_0 \, b_0''  \end{array}\right)
  \nonumber \\
  && + \left(\begin{array}{cc}
  [ b_1 b_0''-b_0 (b_1''+b_0''')]/b_0^2  & -b_0' \, b_0''    \\
  0 &  (-b_1+b_0') \, b_0''-b_0 \, (b_1''+b_0''') 
  \end{array}\right) \, \ell + {\cal O}(\ell^2)   \qquad  \\
   \overline{\rm Ric} &=&\left(\begin{array}{cc}
   b_0''/b_0 & 0 \\ 0 & b_0 \, b_0'' \end{array}\right) \nonumber \\
   && + \left(\begin{array}{cc}
    -(b_1+b_0') \, b_0''/b_0^2 + (b_1''+b_0'')/b_0  & - b_0' \, b_0''   \\
   0 &  b_1 \, b_0'' + b_0 \, (b_1''+b_0''') 
   \end{array}\right)\, \ell  + {\cal O}(\ell^2)     \qquad  \\
  \widetilde{\rm Ric} &=& \left[-{b_0'' \over b_0}+{2 b_1 b_0''+ 
   b_0' b_0''-2b_0 (b_1''+b_0''') \over 2 b_0^2} + 
   {\cal O}(\ell^2) \right] \,  g_0 
\ee
where $g_0 := \mbox{diag}(1,b_0^2)$. For the curvature scalar 
we obtain
\be
   \hat{R} = -{2b_0'' \over b_0}+{2 b_1 b_0''+ 
   b_0' b_0''-2b_0(b_1''+b_0''') \over  b_0^2} + 
   {\cal O}(\ell^2)
\ee

\begin{figure} 
\centering
\includegraphics{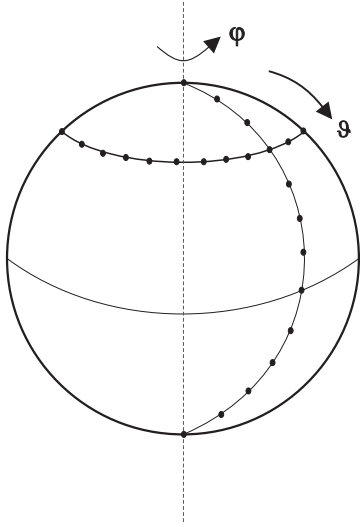} 
\caption{\small Discretization of a sphere.} 
\end{figure}
 
\vskip.2cm
\noindent
{\em Example.}
In ordinary continuum differential geometry, the standard geometry of the unit 
sphere is obtained with $b(\th) = \sin \th$. With this choice,  we get 
\be
    \hat{R} = 2 + \ell \, \mbox{cot} \, \th + {\cal O}(\ell^2)  \; .
\ee
in the discrete framework and in the limit $\ell \to 0$ we recover the continuum 
result $\hat{R} = 2$. To first order, there is a dependence
of the curvature scalar on $\th$. With the refined choice 
$b(\th,\ell) = [1+\th\ell/4+{\cal O}(\ell^2)] \, \sin \th$, we get 
\be
     \hat{R} = 2 + {\cal O}(\ell^2)  \; .
\ee
                             \hfill $\blacksquare$
\vskip.2cm

Our discrete version of curvature describes {\em finite} distances on a space
in contrast to infinitesimal distances as expressed by tangent vectors in continuum 
differential geometry. This means that the metric components in the case under
consideration have to be expected to depend on the discretization (which should be
regarded as a discretization of a chart), i.e., on $\ell$ in the case under consideration. 
We still have to understand how, for example, spherical symmetry can be formulated
in our framework. Then, we should be able to determine a spherically symmetric
metric as a suitable discrete counterpart of the Riemannian metric of the (continuum)
sphere. Furthermore, it remains to be seen how this is related to the metric with 
constant curvature scalar, approximated in the above example.

\section{Conclusions}
\setcounter{equation}{0}
Within a framework of noncommutative geometry, we have presented a formalism 
of discrete Riemannian geometry which is very much analogous to continuum
Riemannian geometry.
\vskip.2cm

Whereas the general formalism of noncommutative geometry suggests to consider
a (generalized) metric tensor as an element of $\O^1 \oa \O^1$, in this paper 
it was taken to be an element of $\O^1 \otimes_L \O^1$ since
a simple geometric meaning can be assigned to its components (with respect to 
the canonical basis $e^{ij}$ of $\O^1$, cf section 3).\footnote{In the case of
a commutative algebra $\A$, one can think of replacing more generally $\oa$
by $\otimes_L$ in  basic definitions like that of a connection. For a noncommutative 
differential calculus, this turns out to be inconsistent with the Leibniz rule, however.
Also, it should be clear that the connection must be a non-local object, in contrast 
to something like a metric tensor.}  
\vskip.2cm

The compatibility condition $\nabla g = 0$ for a metric and a linear connection on a 
finite set, when expressed in terms of parallel transport matrices, 
leads to relations (cf section 3.5) which are in complete accordance 
with what one should expect on the basis of a reformulation of metric compatibility 
in terms of parallel transport in (continuum) differential geometry.
\vskip.2cm

An important role in ordinary differential geometry and especially in General Relativity
is played by the Ricci tensor and the curvature scalar. There is no
generalization of these tensors to the general framework of noncommutative
geometry. In the case of a discrete set, we considered this problem in some
detail in section 4 and showed that, with certain restrictions on the
differential calculus (and thus the links between the points of the set),
satisfactory candidates for discrete counterparts of the continuum Ricci tensor and
curvature scalar do exist. The examples treated in sections 4-6 demonstrate how
our definitions work. It should be quite evident by now that general
definitions can hardly be expected since in noncommutative geometry, and 
already with a commutative algebra $\A$, we are dealing with a huge variety
of structures of which only few should be expected to be close (in some sense)
to continuum differential geometry.

\vskip.2cm
In the last two sections we have developed discrete differential geometry on a 
hypercubic lattice. Since we were able to construct a Ricci tensor and a curvature
scalar in this case, discrete counterparts of the (vacuum) Einstein equations are 
obtained. The results of the last section suggest to choose the following version,
\be                    
  \widetilde{\rm Ric}_{\mu \nu} - {1 \over 2} \, \widetilde{R} \, g_{\mu \nu} 
   = 0   \; .
\ee
On the left hand side we have tensor components in the sense that they transform
covariantly under a change of module basis in the space of 1-forms.
It is straightforward to include matter fields in this scheme. The `discrete gravity'
theory which we propose here is very different from earlier approaches which
were either based on Regge calculus \cite{Regg61}, other simplicial complex
structures \cite{simpl_compl},  or on a certain reformulation of gravity as a gauge 
theory \cite{latt_gravity}. 
The correspondence between first order differential calculi on discrete sets and 
digraphs relates our formalism to the spin network approach to (quantum) gravity 
(see \cite{spinnet}, in particular) at least on a basic level.

\vskip.3cm
\noindent
{\bf Acknowledgments.} A D is grateful to Professor Theo Geisel for financial
support during a stay at the Max-Planck-Institut f\"ur Str\"omungsforschung.
 F M-H would like to thank John Madore for a discussion.


\begin{thebibliography}{99}

\bibitem{DMH92_qlatt} A. Dimakis and F. M\"uller-Hoissen,
  ``Quantum mechanics on a lattice and $q$-deformations'',
 Phys. Lett. B {\bf 295}, 242 (1992).
\bibitem{DMH94_fin} A. Dimakis and F. M\"uller-Hoissen,
 ``Differential calculus and gauge theory on finite sets'',
  J. Phys. A {\bf 27}, 3159 (1994).
\bibitem{DMH94_ddm} A. Dimakis and F. M\"uller-Hoissen,
 ``Discrete differential calculus, graphs, topologies and gauge theory'',
  J. Math. Phys. {\bf 35}, 6703 (1994).
\bibitem{DMHV95} A. Dimakis, F. M\"uller-Hoissen and F. Vanderseypen,
 ``Discrete differential manifolds and dynamics on networks'',
  J. Math. Phys. {\bf 36}, 3771 (1995).
\bibitem{BDMHS96} K. Bresser, A. Dimakis, F. M\"uller-Hoissen and A. Sitarz, 
 ``Non-commutative geometry of finite groups'',
 J. Phys. A {\bf 29}, 2705 (1996).
\bibitem{DMHS93} A. Dimakis, F. M\"uller-Hoissen and T. Striker,  
 ``Noncommutative differential calculus and lattice gauge theory'',
  J. Phys. A {\bf 26}, 1927 (1993).
\bibitem{DMH96_id} A. Dimakis and F. M\"uller-Hoissen,  
 ``Integrable  discretizations of chiral models via deformation of the differential
 calculus'', J. Phys. A {\bf 29}, 5007 (1996).   
\bibitem{Conn94} A. Connes,  {\em Noncommutative Geometry} 
 (Academic Press, San Diego, 1994).
\bibitem{DMH97_dist} A. Dimakis and F. M\"uller-Hoissen,  
 ``Connes' distance function on one-dimensional lattices'',
 Int. J. Theor. Phys. {\bf 37}, 907 (1998).
\bibitem{Sita94} A. Sitarz, ``Gravity from noncommutative geometry'',
 Class. Quantum Grav. {\bf 11}, 2127 (1994); 
 ``On some aspects of linear connections in noncommutative geometry'', 
 preprint hep-th/9503103.
\bibitem{DMMM95} M. Dubois-Violette, J. Madore, T. Masson and J. Mourad,
 ``Linear connections on the quantum plane'', Lett. Math. Phys.
 {\bf 35}, 351 (1995).
\bibitem{Heck+Schm95} I. Heckenberger and K. Schm\"udgen, ``Levi-Civita
 connections on the quantum groups $SL_q(N)$, $O_q(N)$ and $Sp_q(N)$'',
 q-alg/9512001.
\bibitem{Dima96} A. Dimakis, ``A note on connections and bimodules'', 
 q-alg/9603001.
\bibitem{DMH96_deform} A. Dimakis and F. M\"uller-Hoissen,
 ``Deformations of classical geometries and integrable systems'', 
 preprint physics/9712002.
\bibitem{Cho+Park97} S. Cho and K.S. Park, ``Linear connections on graphs'',
 J. Math. Phys. {\bf 38}, 5889 (1997).
\bibitem{Zapa97} R.R. Zapatrin, ``Polyhedral representations of discrete differential
 manifolds'', J. Math. Phys. {\bf 38}, 2741 (1997).
\bibitem{BDMH95} H.C. Baehr, A. Dimakis and F. M\"uller-Hoissen, 
 ``Differential calculi on commutative algebras'',  J. Phys. A {\bf 28},
  3197 (1995). 
\bibitem{Brew97} L. Brewin, ``Riemann normal coordinates, smooth 
 lattices and numerical relativity'', gr-qc/9701057.
\bibitem{Regg61} T. Regge, ``General relativity without coordinates'',
 Nuovo Cim. A {\bf 19}, 558 (1961);
 R.M. Williams and P.A. Tuckey, ``Regge calculus: a brief review 
 and bibliography'', Class. Quantum Grav. {\bf 9}, 1409 (1992).
\bibitem{simpl_compl} D. Weingarten, ``Geometric formulation of
 electrodynamics and general relativity in discrete space-time'', J. Math. Phys.
 {\bf 18}, 165 (1977); \\
 A. N. Jourjine, ``Discrete gravity without coordinates'', 
  Phys. Rev. {\bf  D35}, 2983 (1987).
\bibitem{latt_gravity}
 A. Das, M. Kaku and P.K. Townsend, ``Lattice
 formulation of general relativity'', Phys. Lett. B {\bf 81}, 11  (1979); \\
 L. Smolin, ``Quantum gravity on a lattice'', Nucl. Phys. 
 B {\bf 148}, 333 (1979);  \\
 K.I. Macrea, ``Rotationally invariant field theory on lattices.
 III. Quantizing gravity by means of lattices'', Phys. Rev. {\bf D23}
 900 (1981); \\
 C.L.T. Mannion and J.G. Taylor, ``General relativity on a
 flat lattice'', Phys. Lett. B {\bf 100}, 261 (1981);   \\
 K. Kondo, ``Euclidean quantum gravity on a flat lattice'', Progr. Theor. Phys.
 {\bf 72}, 841 (1984); \\
 M. Caselle, A. D'Adda and L. Magnea, ``Lattice gravity and supergravity
 as spontaneously broken gauge theories of the (super) Poincar{\'e} group'',
 Phys. Lett. B {\bf 192}, 406 (1987), ``Doubling of all matter fields coupled to
 gravity on a lattice'', Phys. Lett. B {\bf 192}, 411 (1987); \\
 P. Renteln and L. Smolin, ``A lattice approach to spinorial quantum gravity'',
 Class. Quantum Grav. {\bf 6}, 275 (1989); \\ 
 O. Bostr\"om, M. Miller and L. Smolin, ``A new discretization of
 classical and quantum general relativity'', preprint CGPG 94-3-3; \\
 R. Loll, ``Discrete approaches to quantum gravity in four dimensions'',
 gr-qc/9805049.
\bibitem{spinnet} A. Ashtekar and J. Lewandowski, ``Quantum theory of
 geometry I: area operators'', Class. Quantum Grav. {\bf 14}, A55 (1997); \\
 F. Markopoulou and L. Smolin, ``Causal evolution of spin networks'', 
 Nucl.Phys. B {\bf 508}, 409 (1997); \\
 M.P. Reisenberger, ``A lattice worldsheet sum for 4-d Euclidean general
 relativity'', gr-qc/9711052.
\end{thebibliography}
\end{document}